\documentclass[]{aa}   

\usepackage{color}
\definecolor{black}{rgb}{0,0,0}
\definecolor{red}{rgb}{1,0,0}
\definecolor{darkblue}{rgb}{0,0,0.7}
\definecolor{blue}{rgb}{0,0,1} 
\definecolor{green}{rgb}{0,0.5,0} 
\definecolor{orange}{rgb}{0.8,0.6,0} 
\definecolor{purple}{rgb}{1,0,1}

\usepackage{orcidlink}
\usepackage{longtable,array}
\usepackage{graphicx}
\usepackage{txfonts}
\usepackage[utf8]{inputenc} 
\usepackage{float}

\begin{document}

   \title{
     Rings around irregular bodies.}

   \subtitle{II. Numerical simulations of the 1/3 spin-orbit resonance confinement and applications  to Chariklo}

\author{
H. Salo$^{\orcidlink{0000-0002-4400-042X}}$\inst{1} %
\and
B. Sicardy$^{\orcidlink{0000-0003-1995-0842}}$\inst{2} 
}

\institute{
Space Physics and Astronomy Research unit, University of Oulu, FI-90014 Oulu, Finland\\
\email{heikki.salo@oulu.fi}
\and
Laboratoire Temps Espace (LTE), Observatoire de Paris, Universit\'e PSL, CNRS UMR 8255, Sorbonne Universit\'e, LNE, 61 Av. de l'Observatoire, F75014 Paris, France
}
 
\date{Received August 21, 2025; accepted January 3, 2026}


  \abstract
   {%
    Rings have been found around Chariklo, Haumea and Quaoar, three small objects of the Solar System. All these rings are observed near the second-order spin-orbit resonances (SORs) 1/3 or 5/7 with the central body, suggesting an active confinement mechanism by these resonances.
   }%
   {%
   Our goal is to understand how collisional rings can be confined near second-order SORs in spite of the fact that they force self-intersecting streamlines.
   }%
   {%
   We use full 3D numerical simulations that treat rings of inelastically colliding particles orbiting non-axisymmetric central bodies, characterized  by a dimensionless mass anomaly parameter $\mu$.  
    While most of our simulations ignore self-gravity, a few runs include gravitational interactions between particles, providing preliminary results on the effect of self-gravity on the ring confinement. 
   }%
   {%
    The 1/3 SOR can confine ring material, by transferring the forced resonant mode into free Lindblad modes. We derive a criterion ensuring that the 1/3 SOR counteracts viscous spreading. It reads $k \mu^2 \gtrsim \tau R^2$, where $k$ is a dimensionless coefficient, $\tau$ is the ring optical depth and $R$ is the particle radius. Expressing $R$ in terms of the radius of the synchronous orbit, we obtain $k \sim 4 \times 10^{-5}$ for the 1/3 SOR acting on non-gravitating rings. Assuming meter-sized ring particles, and $\tau \sim1$, this requires a threshold value $\mu \gtrsim 10^{-3}$ in Chariklo's case.
   The confinement is not permanent as a slow outward leakage of particles is observed in our simulations. This leakage can be halted by an outside moonlet with a mass of $\sim 10^{-7}-10^{-6}$ relative to Chariklo, corresponding to subkilometer-sized objects.  With self-gravity, the ring viscosity increases by a factor of few in low-$\tau$ rings due to gravitational encounters. For large $\tau$, self-gravity wakes enhance the viscosity $\nu$ by a factor of $\sim$100 compared to a non-gravitating ring, requiring $\sim$10-fold larger $\mu$'s since the threshold value increases proportional to $\sqrt{\nu}$. 
   }%
   {}
   
   \keywords{
    Celestial mechanics ---
    Planets and satellites: rings
    }

\titlerunning{Numerical simulations of the 1/3 spin-orbit confinement of Chariklo's ring}
   \maketitle

   \nolinenumbers
   
\section{Introduction}

Since 2013, narrow and dense rings have been discovered around four small objects of the outer Solar System:
the Centaur Chariklo \citep{braga2014}, the dwarf planet Haumea \citep{ortiz2017}, the large trans-Neptunian object Quaoar \citep{morgado2023,pereira2023}, and possibly the Centaur Chiron \citep{ortiz2023}, see also the reviews by \cite{sicardy2024b} and \cite{sicardy2025a}. 
Current observations show that Chariklo's and Haumea's rings are close to the second-order 1/3 spin orbit resonance (SOR) with the central body, meaning that the ring particles complete one revolution while the central body completes three rotations. The two rings of Quaoar are also near second-order resonance, the main ring being again near the 1/3 SOR, while the fainter ring orbits near the 5/7 SOR. 

In \cite{sicardy2025b} (Paper~I hereafter), we have studied the topological structure of the phase portraits corresponding to resonances of the first to fifth order. It appears from this work that among all these resonances, only those of order one (aka Lindblad) and two are expected to significantly perturb a collisional ring where eccentricities are damped by dissipative impacts. The reason for this is that only for these two kinds of resonances the origin of the phase portrait (corresponds to zero eccentricity) is not a stable (elliptical) fixed point. 

This paper is the numerical counterpart of Paper~I, first to validate the special status of the first- and second-order SORs mentioned above, and second to explore the ability of the 1/3 SOR to confine collisional rings. As mentioned in Paper~I, only the first-order resonances force streamlines that do not self-intersect. As such, they can create non-singular spiral waves that have been studied for decades in the galactic and planetary ring contexts.
In contrast, any periodic orbit forced by a resonance of order greater than one has at least one self-intersecting point, thus preventing in principle a regular response from a collisional disk. This is true in particular with the 1/3 SOR that forces streamlines with one self-intersecting point.

 Here we explore using numerical simulations how the 1/3 SOR can confine material, in spite of the self-intersection problem.  Compared to existing N-body simulation studies for small-body rings, our models include simultaneously all three ingredients necessary for a realistic modeling of SOR resonances: i) a rotating non-axisymmetric central body,
ii) particle physical collisions, and iii) an azimuthally complete 3D ring. For example, \cite{michikoshi2017} performed large-N body collisional simulations using an azimuthally complete ring, but assumed a spherical central body. \cite{gupta2018} made collisional simulations with a non-spherical potential, but since the simulations were done in a local co-moving frame, they were limited to a case of
axisymmetric non-rotating central potential. The two studies mentioned above thus missed SOR resonance effects.

On the other hand, \cite{sickafoose2024} suggested that Chariklo's rings could be stabilized by a Lindblad resonance with a nearby shepherd satellite. Their simulations used a modified local code, where the semi-periodic azimuthal boundary conditions were designed to correspond to the streamlines of the assigned first order resonance. It is thus not obvious how to apply such a method to 1/3 SOR or to higher order resonances in general. The mechanism itself, the collisional confinement of rings at first order resonances, was already found in simulations of \cite{hanninen1994, hanninen1995}, which followed azimuthally complete rings. In the current study we demonstrate how similar confinement takes place in rings perturbed by second order resonances relevant for the observed systems.

This paper is organized as follows. In Section 2 we summarize our simulation method, while Section 3 reviews results of non-collisional test-particle simulations and their good agreement with the analytical calculations of Paper~I. In Sections 4-7 collisional simulations are reported: after providing an illustration of the 1/3 SOR confinement, we establish a scaling of the simulation results to real systems and estimate the sizes of mass anomaly capable of confining ring material around small irregular bodies (Sect. 4), explore the normal modes excited inside confined ringlets (Sect. 5), and the outward migration of ring material through various resonances (Sect. 6). In Section 7 we demonstrate that the long term stabilization of the ringlet is possible via a torque from a small additional external satellite. However, even in this case the location and confinement of the ringlet is determined by the 1/3 SOR. Finally, Section 8 shows that the confinement mechanism works also in self-gravitating rings.  This paper describes in more detail the preliminary results obtained by \cite{salo2021,sicardy2021,salo2024} and
\cite{sicardy2024a}.

\section{Simulation method}

Our numerical simulations treat a ring of inelastically colliding particles around a non-axisymmetric central body rotating at angular speed $\Omega_{\rm B}$. Several models for the potential of the central body can be used, see details in Appendix~\ref{app_potentials}:
(i) a homogeneous triaxial ellipsoid, and
(ii) a spherical body with a dimensionless mass anomaly $\mu$ at its surface. These two models can also be combined so that
(iii) the mass anomaly is attached to the triaxial ellipsoid. Finally, 
(iv) the ring particles can be perturbed by an additional satellite orbiting the central body.  

Calculations extending over tens of thousands of central body revolutions are performed with IDL (Interactive Data Language) using a RK4 integrator in double precision. Impacts between ring particles are treated as soft-sphere collisions by including visco-elastic pressure forces between colliding, slightly penetrating particles. The treatment of collisions follows the schemes presented by \cite{salo1995}, except that azimuthally complete rings are now simulated instead of local co-moving ring patches. In all our simulations, the adopted coefficient of restitution is $\epsilon_{\rm n} = 0.1$, see more details in Appendix~\ref{app_treatment_impacts}.  Most of our simulations ignore mutual gravity between particles. However, Sect. \ref{sec_sg} reports preliminary simulations with particle-particle calculation of ring self-gravity. 

In this paper, $a$, $e$, $L$ and $\varpi$ denote the osculating orbital semimajor axis, eccentricity, true longitude and longitude of pericenter of the particles, respectively, as calculated in the center-of-mass reference frame.  The Appendix~\ref{app_potentials} also provides the definitions of the reference radius $R_{\rm ref}$, the elongation $\epsilon_{\rm elon}$ and the oblateness $f$ of a homogeneous triaxial ellipsoid, used later in this paper.

\section{Results of test particle integrations}

\subsection{Triaxial central body vs mass anomaly}
\label{sec_triaxial_vs_mass_anom}

Spin-orbit resonances (SORs) occur near commensurabilities between $\Omega_{\rm B}$ and the mean motion $n$ of the ring particles, see Paper~I. One kind of SOR is the corotation resonance that occurs for $n=\Omega_{\rm B}$, i.e. at the synchronous orbit with radius $a_{\rm cor}$. This resonance will not be studied here because the corresponding equilibrium points (akin to the Lagrange triangular points L$_4$ and L$_5$) are dynamically unstable or close to instability in the cases of Chariklo and Haumea \citep{sicardy2019}. Moreover, they are also generally unstable against dissipative collisions since they correspond to local maxima of potential.

The other SORs occur for $j \kappa= m(n-\Omega_{\rm B})$, where $\kappa$ is the particle epicyclic frequency, $m$ is the azimuthal number of the resonance (positive or negative) and $j>0$ is its order. These SORs excite the orbital eccentricities of the particles and, as shown in this paper, can confine rings in narrow regions. If the particles' precession rate $\dot{\varpi}= n -\kappa$ is small compared to $n$, the above condition reads
\begin{equation}
  \frac{n}{\Omega_{\rm B}} \approx \frac{m}{m-j}.
  \label{eq_ratio_n_OmegaB}
\end{equation}
Following the notation of Paper~I, $m \ge 1$ corresponds to inner resonances ($n > \Omega_{\rm B}$), while $m \le -1$ corresponds to outer resonances $(n < \Omega_{\rm B})$. As mentioned in the Introduction, only resonances with $j=1$ and $j=2$ are able to excite the orbital eccentricities of particles colliding inelastically. Following Paper~I, we define the quantity 
\begin{equation}
 \overline{a} = a + a_0 \left( \frac{m-j}{j} \right) e^2
 \label{eq_modified_a}
\end{equation}
as the ``modified semimajor axis". It is a local expression of the Jacobi constant for a given $m/(m-j)$ SOR, where $a_0$ is the semimajor axis of the circular orbit at exact resonance.  The modified semimajor axis is essentially a measure of the semimajor axis $a$ for $e \ll 1$. The advantage of $\overline{a}$ over $a$ is that it is on the average constant near a resonance, so that particles trapped into this resonance will tend to move vertically in the diagrams showing the eccentricity plotted vs. $\overline{a}$.

The results given in this paper are applicable to any ring around any body with a given mass anomaly $\mu$ or triaxial shape (and thus with given reference radius $R_{\rm ref}$, elongation $\epsilon_{\rm elon}$ and oblateness $f$). 

Unless otherwise explicitly stated, times will be expressed in terms of the number of rotations of the central body (corresponding to $2\pi$ time units) and the lengths (including the particle radius $R$) will be expressed in units of the corotation orbit radius $a_{\rm cor}$. The mass anomaly, if present, is located at the distance $R_{\rm ref}$.
To relate these quantities to more physical cases, we can consider for instance the case of Chariklo, using the parameters given in Table~\ref{tab_param_cha}. The cases of Haumea and Quaoar can also be considered, using the values provided in the  Table~2  of Paper~I.

\begin{table}[!h]
\caption{Adopted physical parameters of Chariklo.
\label{tab_param_cha}}
\begin{tabular}{ll}
\hline \hline
Mass $M$                          & $7 \times 10^{18}$ kg         \\
Semiaxes $A \times B \times C$   & $157 \times 139 \times 86$ km \\
Reference radius $R_{\rm ref}$    & 115 km  \\
Elongation $\epsilon_{\rm elon}$  & 0.20    \\
Oblateness $f$                    & 0.55    \\
Rotation period                   & 7.004 h \\
Corotation radius                 & 196  km \\
\hline
\end{tabular}
\tablefoot{ Numerical values the same as in Paper I, from \cite{leiva2017, Morgado2021}}
\end{table}

\begin{figure}[!t]
  \centering
          \includegraphics[width=0.95\columnwidth]{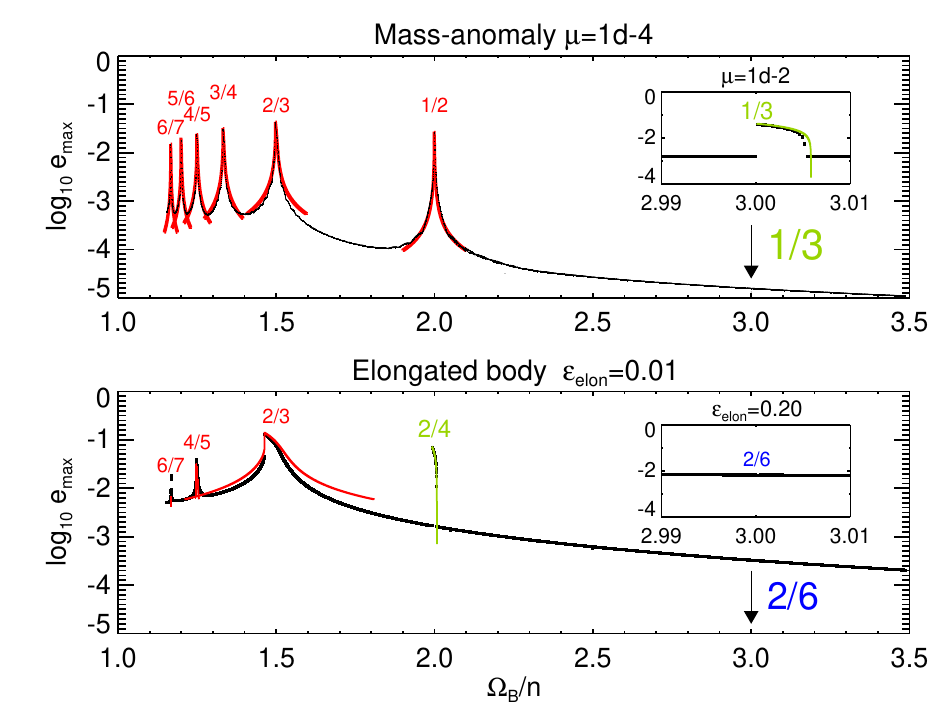}
          \caption{
            The comparison between the numerical and theoretical responses of test particles to various SORs.
            Numerical integrations have followed the motion of 10,000 test particles initially distributed 
            on circular orbits, during 10,000 revolutions of the central body. The maximum eccentricities $e_{\rm max}$ 
            reached by these particles are plotted in black as a function of $\Omega_{\rm B}/n =(a/a_{\rm cor})^{3/2}$, 
            and are compared with the analytical estimates of Paper~I (red or green curves). 
            \textit{Upper panel}\rm: the case of a spherical body with a mass anomaly $\mu=10^{-4}$. 
            \textit{Lower panel}\rm: the case of a homogeneous triaxial ellipsoid with elongation
            $\epsilon_{\rm elon}=0.01$ and oblateness $f=0$.
            In the case of mass anomaly the strongest responses are at the outer
            first-order Lindblad resonances corresponding to
            commensurabilities $n/\Omega_{\rm B}= m/(m-1)$, with $m=-1,-2,-3$... 
            The insert in the upper panel is a zoom on the $\Omega_{\rm B}/n= 3$ region, 
            using a 100 times larger mass anomaly of $\mu=0.01$. The response to the second-order 1/3 resonance 
            is now visible and compared to the analytical estimate in green. 
            In a case of an elongated body, the $\pi$-symmetry of the perturbation imposes
            even values $m=-2,-4,-6$... for the first-order resonances. 
            In this case the strongest second-order resonance 
            ($m=-2$ and $j=2$) is also visible, with the analytical response plotted in green. 
            The insert in the lower panel is a zoom on the $\Omega_{\rm B}/n= 3$ region: the 
            fourth-order 2/6 resonance has no signature even when using a large value $\epsilon_{\rm elon}=0.20$.
            }
          \label{fig_num_vs_theo_e}
      \end{figure}

Examples of test particle responses to various outer SORs are displayed in Fig.~\ref{fig_num_vs_theo_e}, in terms of the maximum eccentricity $e_{\rm max}$ achieved by particles initially on circular orbits. Orbits around a spherical central body with a mass anomaly $\mu=10^{-4}$ (upper frame) and an ellipsoidal body with elongation $\epsilon_{\rm elon}=10^{-2}$ and oblateness $f=0$ (lower frame) were integrated for 10,000 revolutions of the central body.

In case of mass anomaly, the perturbation potential includes a full range of $m$-components, leading to a strong response at first-order resonances where  $n/\Omega_{\rm B} =1/2, 2/3, 3/4...$, corresponding to $m=-1,-2,-3...$ with $j=1$. Even with a mass anomaly as small as $\mu=10^{-4}$, test particles reach orbital eccentricities as high as $e_{\rm max}=$~0.01-0.05. Meanwhile, due to the $\pi$-symmetry of the ellipsoidal potential, only even $m$ components are allowed. Consequently, the only first-order resonances present are those corresponding to $n/\Omega_{\rm B} =2/3, 4/5,...$. Thus, the response at $n/\Omega_{\rm B}=1/2$ for an elongated body, visible in Fig.~\ref{fig_num_vs_theo_e}, actually corresponds to a second-order resonance with $m=-2, j=2$, and is thus noted 2/4. Also shown in this figure are the analytical responses calculated in Paper~I, showing a good agreement with our numerical integrations.

We will focus in this paper on the 1/3 SOR, near which the main rings of Chariklo, Haumea and Quaoar are observed \citep{sicardy2025a}, noting that Quaoar's fainter ring is close to another SOR with $n/\Omega_{\rm B} \approx 5/7$ \citep{pereira2023}. In the case of a mass anomaly, the 1/3 SOR corresponds to a second-order resonance with $m=-1,j=2$. Conversely, for an ellipsoidal body and its associated $\pi$-symmetry, the 1/3 SOR corresponds to a fourth-order SOR with $m=-2,j=4$, thus noted 2/6. Concerning the 5/7 SOR, it can be created only by a mass anomaly, thus corresponding to a second-order SOR with $m=-5,j=2$.

With a mass anomaly $\mu=10^{-4}$, no visible signature is seen in Fig.~\ref{fig_num_vs_theo_e} at the 1/3 SOR location. Increasing $\mu$ to $10^{-2}$ (insert in the upper panel of Fig.~\ref{fig_num_vs_theo_e}), a clear response to the 1/3 SOR is seen, in agreement with the theoretical expectation.  On the other hand, for an elongated body, no response is noticeable near the 2/6 SOR, even with an elongation as high as $\epsilon_{\rm elon}=0.2$, corresponding to the shape of Chariklo. This  absence of response is as expected for a fourth-order resonance: as shown in  Fig.~1 of  Paper~I based on the topology of phase portraits, only first and second order resonances excite eccentricities of particles initially  on  circular orbits. 

From hereon we concentrate on first and second-order SOR resonances with a mass anomaly.

\subsection{Scaling of first and second-order resonances}

\begin{figure}
  \centering
        \includegraphics[width=\columnwidth]{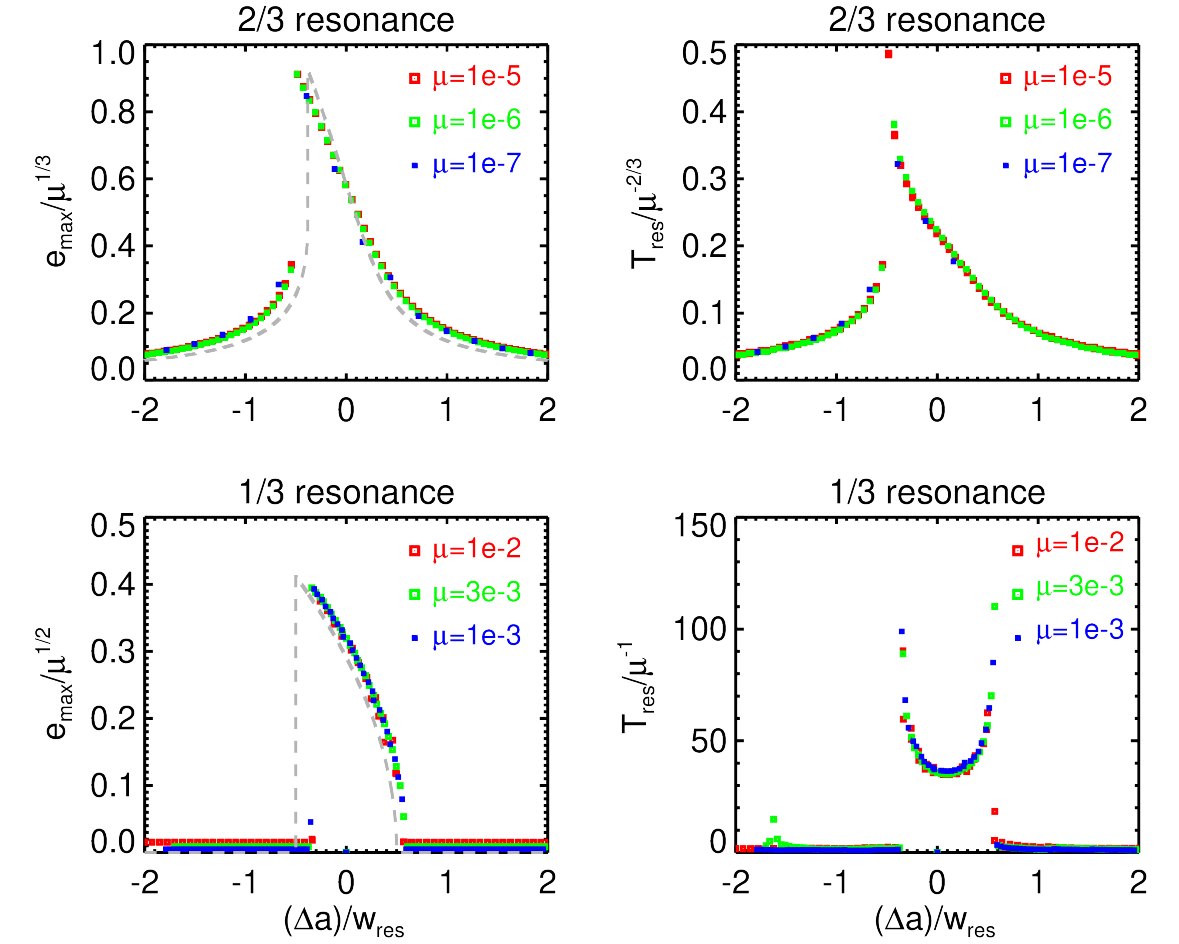}
        \caption{
          The responses of test particles to the first-order 2/3 and second-order 1/3 SORs.
          \textit{Left column}: the symbols represent the maximum eccentricities $e_{\rm max}$ reached by particles initially 
          on circular orbits for three values of the  mass anomaly $\mu$, near the 2/3 SOR (upper panel) and the 1/3 SOR (lower panel).
          The gray dashed curves are the analytical estimates of $e_{\rm max}$  displayed in Figs.~3 and 4 of  Paper~I.
          The points are plotted against the normalized distance to the resonance, $\Delta a/W_{\rm res}= (\overline{a} - a_0)/W_{\rm res}$, where 
          $a_0$ is the radius of the circular orbit at exact resonance, $\overline{a}$ is the modified semimajor axis (Eq.~\ref{eq_modified_a}) and $W_{\rm res}$ is the width of the resonance (Eq.~\ref{eq_peak}).
          \textit{Right column}: 
          the same with the timescales $T_{\rm res}$ necessary to reach the maximum eccentricities $e_{\rm max}$.
          This figure shows that our numerical integrations reproduce satisfactorily the calculated distribution of $e_{\rm max}$,
          as well as the $\mu$-scaling expected from Eqs.~\ref{eq_peak} and \ref{eq_tres}.
          }
          \label{fig_num_vs_theo_scalings}
\end{figure}

We checked that our simulations correctly reproduce the expected scaling of the particle responses against $\mu$.  Figure \ref{fig_num_vs_theo_scalings}  illustrates the responses of test particles near the first-order 2/3 and second-order 1/3 SORs for various $\mu$'s, compared to the values of $e_{\rm max}$  displayed in Figs.~3 and 4 of Paper~I. We denote by $e_{\rm peak}$ the largest possible value of $e_{\rm max}$ near a given resonance and $W_{\rm res}$ is the width the resonance  given in table~1 of  Paper~I. For first-order resonance, $W_{\rm res}$ is close to the FWHM of the $e_{\rm max}$ distribution, while for second-order resonance, it corresponds to the interval where $e_{\rm max}$ is nonzero. In the particular case of Chariklo, and following the methodology presented in Paper~I to derive the resonance strengths, we obtain
\begin{eqnarray}
 &e_{\rm peak}&\!\!\!  \!\approx \!  0.93 \mu^{1/3}, \ W_{\rm res} \approx  \! 2.52\mu^{2/3}  \  ({\rm first-order~2/3~SOR}), \cr
 &e_{\rm peak}&\!\!\!  \!\approx \!  0.41 \mu^{1/2}, \ W_{\rm res} \approx \!  0.26 \mu  \ \  \ ({\rm second-order~1/3~SOR}).
 \label{eq_peak}
\end{eqnarray}
The numerical factors are specific to the Chariklo case, while the $\mu$-scaling depends only on the resonance order.
 
\begin{figure}
  \centering
        \includegraphics[width=0.9\columnwidth]{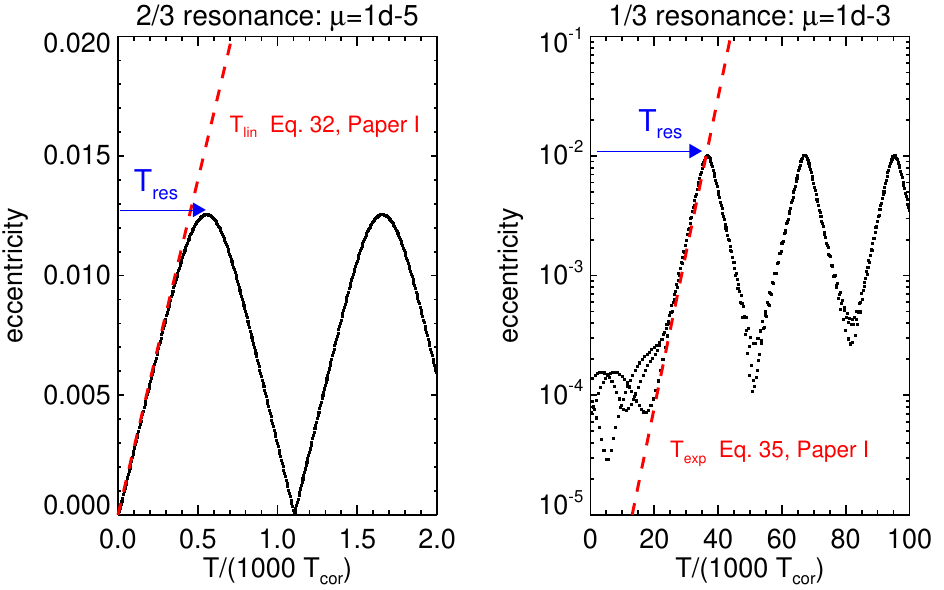}
        \caption{
         Comparison of eccentricity evolution at the first-order 2/3 and second-order 1/3 SORs. The black curves follow the evolution of test particles released near exact resonance, while the red dashed lines display the theoretical prediction, that is a linear growth rate for the first-order SOR and an exponential growth rate for the second-order SOR. 
          }
          \label{fig_num_vs_theo_scalings2}
\end{figure}

The left column of Fig.~\ref{fig_num_vs_theo_scalings} shows that the general shape of the $e_{\rm max}$ distribution, the numerical values of $e_{\rm peak}$ and $W_{\rm res}$ and the $\mu$-scaling are correctly reproduced in our integrations.\footnote{This scaling holds better than 5\% for the $\mu$ range displayed in the figure. For $\mu=0.03$ and $0.1$ the $e_{\rm peak}$ for 1/3 SOR are about 15\% and 30\% smaller than implied by Eq. \ref{eq_peak}; similarly, the times to reach the maximum are somewhat longer than given  by Eq. \ref{eq_tres}.}. The right column shows the time $T_{\rm res}$ required for a particle initially on a circular orbit to reach the maximum value $e_{\rm max}$. Numerical integrations imply that at the resonance 

\begin{eqnarray}
&T_{\rm res}&  \  \approx  0.25\mu^{-2/3} \ \ \ \ \ \ ({\rm first-order~2/3~SOR}),\cr
&T_{\rm res}&   \ \approx  
 40  \mu^{-1} \ \ \ \ \ \ \ \ \ \ \ ({\rm second-order~1/3~SOR}).
\label{eq_tres}
\end{eqnarray}
The above $\mu$ scalings confirm the scaling of timescales obtained in Paper~I. Figure~\ref{fig_num_vs_theo_scalings2}  shows a good agreement between numerical integrations and the analytical estimates for the linear growth rate in the case of first-order resonance ($T_{\rm lin} \propto \mu^{-2/3}$; Eq. 32 in Paper~I), and the exponential growth rate in the second-order resonance ($T_{\rm exp} \propto \mu^{-1}$; Eq. 35 in Paper~I). 

For the typical threshold value $\mu=10^{-3}$ that we will later estimate for confining ring material  (this would correspond to $\sim 10$ km mountain on Chariklo), we obtain at the first-order 2/3 resonance (orbital radius $a_{2/3} = 1.31$) the values $e_{\rm peak}=0.093, W_{\rm res} = 0.0252$ and $T_{\rm res} \approx 600$. For the second-order 1/3 resonance at $a_{1/3} = 2.08$, we obtain $e_{\rm peak} = 0.013, W_{\rm res} = 0.00026$ and $T_{\rm res} \approx$~40,000. The long timescale for the excitation of eccentricities at the 1/3 resonance fully explains the absence of any signature in Fig.~\ref{fig_num_vs_theo_e}. For $\mu=10^{-4}$, even if the maximum eccentricities would be about 0.004, the growth of resonance eccentricities would have required a 40-fold longer length of integration than depicted in Fig.~\ref{fig_num_vs_theo_e}. Moreover, the resonance width ($\approx 3 \times 10^{-5}$) would have been much too small to be discernible in this figure.

\begin{figure}
  \centering
 \includegraphics[width=\columnwidth]{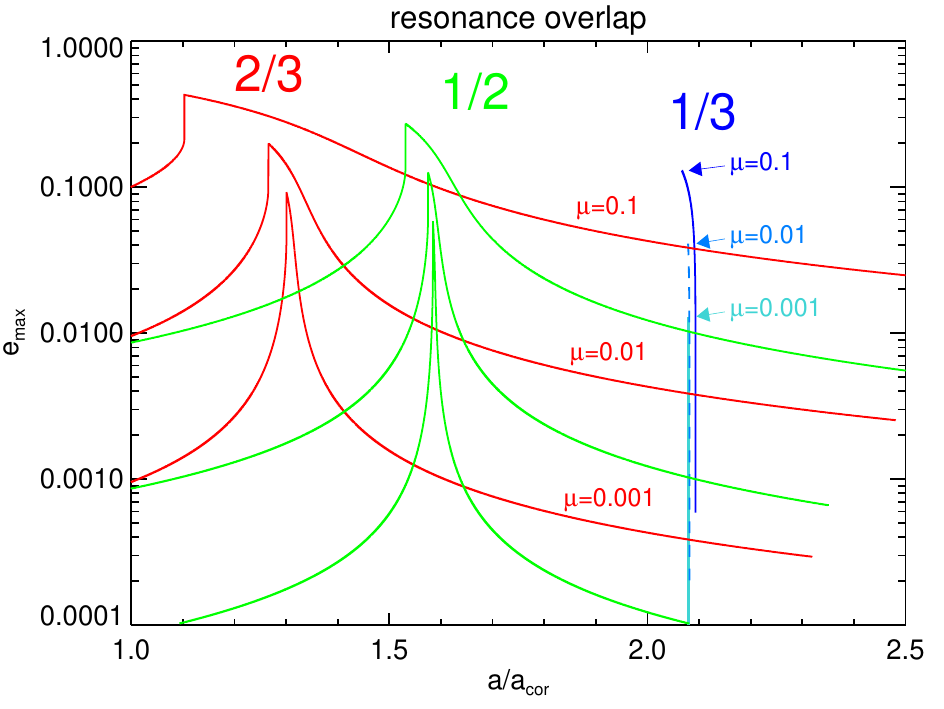}
        \caption{
          Comparison of the maximum eccentricity $e_{\rm max}$ reached by particles at the second-order 1/3 SOR (blue) compared with the first-order 2/3 (red) and 1/2 (green) SORs. Three values $\mu=0.1,0.01,0.001$ are compared. For $\mu=0.1$, the theoretical $e_{\rm max}$ associated with the 2/3 and 1/2 resonances at the location of $1/3$ resonance are $30\%$ and $8\%$ of that
          due 1/3 resonance, respectively. For other $\mu$'s the ratios at the 1/3 SOR  location scale proportional to $\mu^{1/2}$.
        }
        \label{fig_resonance_overlap}
\end{figure}

Because of the large strength and radial extent of the first-order resonances, it is worth estimating what their effect is at the distance of the second-order 1/3 SOR we are interested in. This is illustrated in Fig.~\ref{fig_resonance_overlap}, comparing the theoretical amplitudes $e_{\rm max}$ in the vicinity of 1/3 SOR at $a/a_{\rm cor} \approx 2.08$.  For $\mu=0.1$, the eccentricities associated with the 2/3 SOR at the same distance are $\approx 0.04$, or nearly 30\% of the $e_{\rm peak}$ due to the 1/3 SOR.  Moreover, the maximum eccentricities associated with first-order resonances are reached very rapidly, compared to the slow growth of second-order perturbations. Not surprisingly, collisional simulations performed for 1/3 resonance with large values of $\mu$ also show a clear $m=2$ undulation in their initial evolution phase (for example, see the $T=500$ frames in Fig.\ref{fig_transition_collisionless_collision}). However, the $m=2$ undulation has no effect on the 1/3 resonance confinement observed in collisional simulations\footnote{This was tested in a collisional simulation with $\mu=0.1$ where a potential expansion was used: retaining only the $m=1$ term gave practically identical results to those when using a full mass anomaly potential, while keeping only the $m=2$ term led to no secular effects. See Appendix \ref{AppendixC1}.}. Also, although the values of $e_{\rm peak}$ associated with the first-order resonances decrease more slowly than that of the second-order 1/3 SOR (the ratio of the $e_{\rm peak}$'s scale as $\mu^{-1/6}$, see Eq.~\ref{eq_peak}), the local ratio of eccentricity amplitudes at the 1/3 location drops as $\mu^{1/2}$. This stems from the fact that the widths of the first-order SOR's shrink proportional to $\mu^{2/3}$.

\begin{figure*}[!t]
  \centering
          \includegraphics[width=1.8\columnwidth]
        {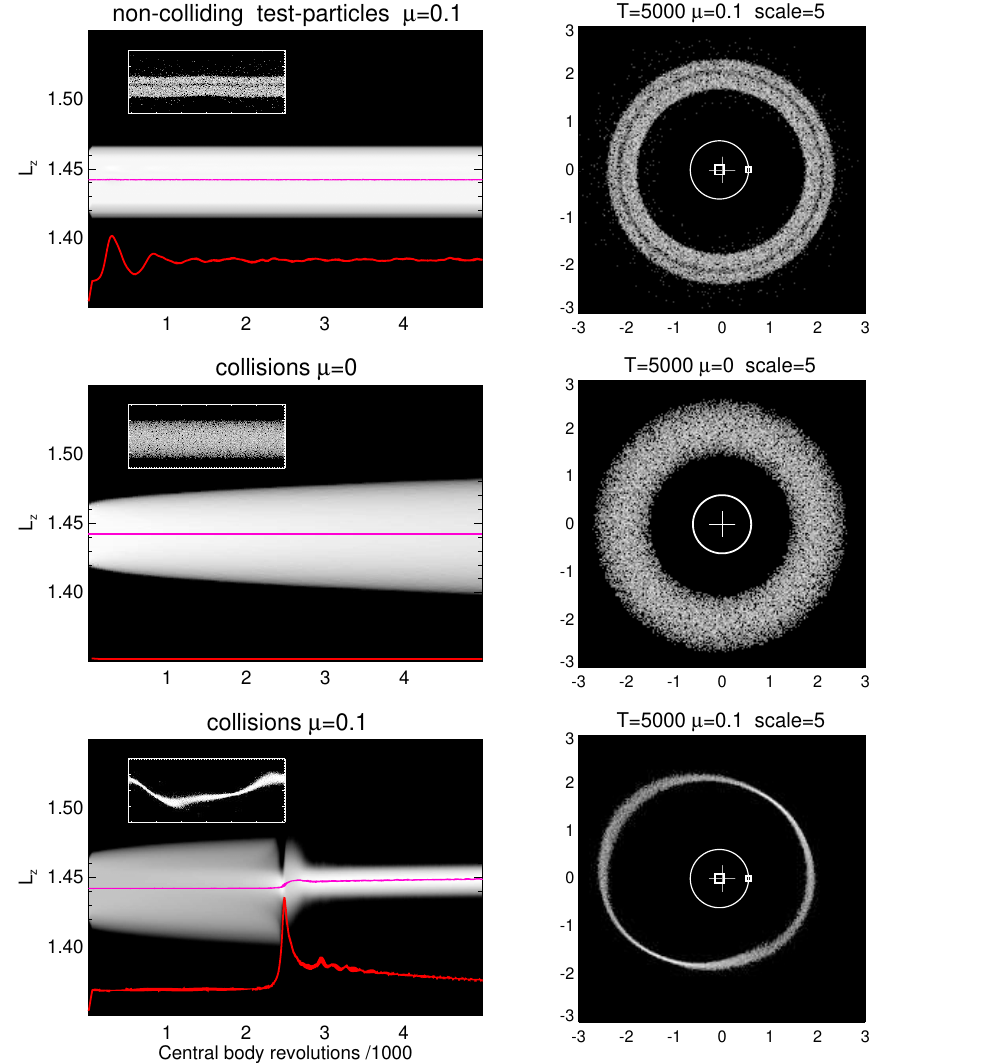}
        \caption{
        Three cases showing the combined effects of collisions and the 1/3 SOR on the ring confinement. The left frames show the time evolution of the vertical angular momentum ($L_z$) distribution of the particles around the 1/3 SOR at $L_z = 1.44$. The magenta lines are the average value of $L_z$ and the red lines show the RMS of the particle eccentricities; the full y-range of the frame corresponds to $e=0.1$. The inserts are polar plots of the particle positions, shown again in the right column in cartesian coordinates. For better viewing, in the cartesian projections both the width of the ring around its center position at each azimuth and the deviations of this center position from the overall mean distance have been expanded by a factor of five. We compare three cases: 
        (i) a ring of collisionless test particles perturbed by a $\mu=0.1$ mass anomaly on an otherwise spherical central body (upper row); 
        (ii) colliding particles around a spherical central body without perturbation ($\mu=0$, middle row); and
        (iii) a ring of colliding particles with $\mu=0.1$ (lower row).
        All simulations use the same initial conditions with 30,000 particles placed initially in an annulus $r=2.02-2.14$ straddling the $1/3$ SOR at semimajor axis $a =2.08$. 
        In the collisional simulations the particle radius $R=10^{-3}$ corresponds to about 200~m for Chariklo's ring particles and yields an initial  geometric optical depth $\tau_0=0.06$. In the case of a mass anomaly, the perturbation is turned on linearly during the first 50 revolutions of the central body.
          }
        \label{fig_combined_effect_collision_sor}
\end{figure*}

\section{Results from collisional simulations}

\subsection{Illustrative example}

We first demonstrate the crucial role of physical impacts on the response to resonances.  This is illustrated in Fig.~\ref{fig_combined_effect_collision_sor} using a large mass anomaly $\mu=0.1$. The particles with radii $R= 10^{-3}$ are initially placed on a wide annulus that straddles the 1/3 resonance semimajor axis $a_{1/3} \approx 2.08$. Recalling that $R$ is measured in units of the corotation radius $a_{\rm cor}$, this corresponds from Table~\ref{tab_param_cha} to a physical radius of $\approx 200$~m for Chariklo's ring particles.

Figure~\ref{fig_combined_effect_collision_sor} shows density plots of the angular momentum distribution $L_z$ evolving with time, together with snapshots of particle positions at the end of the simulation, both in polar and cartesian systems. In the case of non-colliding test particles (upper panels of Fig.~\ref{fig_combined_effect_collision_sor}), the time evolution of the system near the resonance location exhibits a growth of eccentricities, with maximum amplitudes and timescales behaving as illustrated in Fig.~\ref{fig_num_vs_theo_scalings} (see the peak in the eccentricity RMS at $T \approx 400$). Most notably, there is no secular change of particle mean distances, so that there is very little change in the $L_z$ distribution\footnote{Actually, the distribution of Jacobi energy $E_{\rm J}$ (Eq. \ref{eq_ej})  would be a better choice, as the $E_{\rm J}$ of each particle remains constant in the absence of impacts, while $L_z$ is not a constant but oscillates around its mean value. However, the difference is small.}.  When plotting instantaneous particle positions, a gap develops at the resonance due to excited eccentricities, since the particles spend most of their time near the extremes of their orbital epicycles.  

In the case of unperturbed but mutually colliding particles (middle panels of Fig.~\ref{fig_combined_effect_collision_sor}), the ring evolution shows a rapid collisional damping of inclinations and eccentricities, with a timescale of a few tens of impacts per particle, and a gradual viscous spreading taking place on longer timescales determined by the ring viscosity $\nu$. This is seen as a widening of the ring and the $L_z$ distribution in Fig.~\ref{fig_combined_effect_collision_sor}.

Finally, when both impacts and resonant perturbations are included (lower panels of Fig.~\ref{fig_combined_effect_collision_sor}), the behavior of the ring is drastically different. The simulation leads, after initial viscous spreading, to the excitation of eccentricities by the 1/3 SOR in parallel with the accumulation of particles around the resonance. At the same time the mean $L_z$ of the ring jumps and the system settles to a narrow ringlet just outside the resonance. In the simulation of Fig.~\ref{fig_combined_effect_collision_sor} this accumulation and confinement takes place at $T\approx 2500 - 3000$.  We will discuss in Sect.~\ref{sec_scaling} the conditions under which such behavior occurs. However, on longer timescales, not covered in Fig.~\ref{fig_combined_effect_collision_sor}, the ringlet slowly leaks away particles at its outer edge. We return to this process in Sect. \ref{sec_effect_satellite}.

\begin{figure*}[!t]
\centerline{\includegraphics[width=200mm]{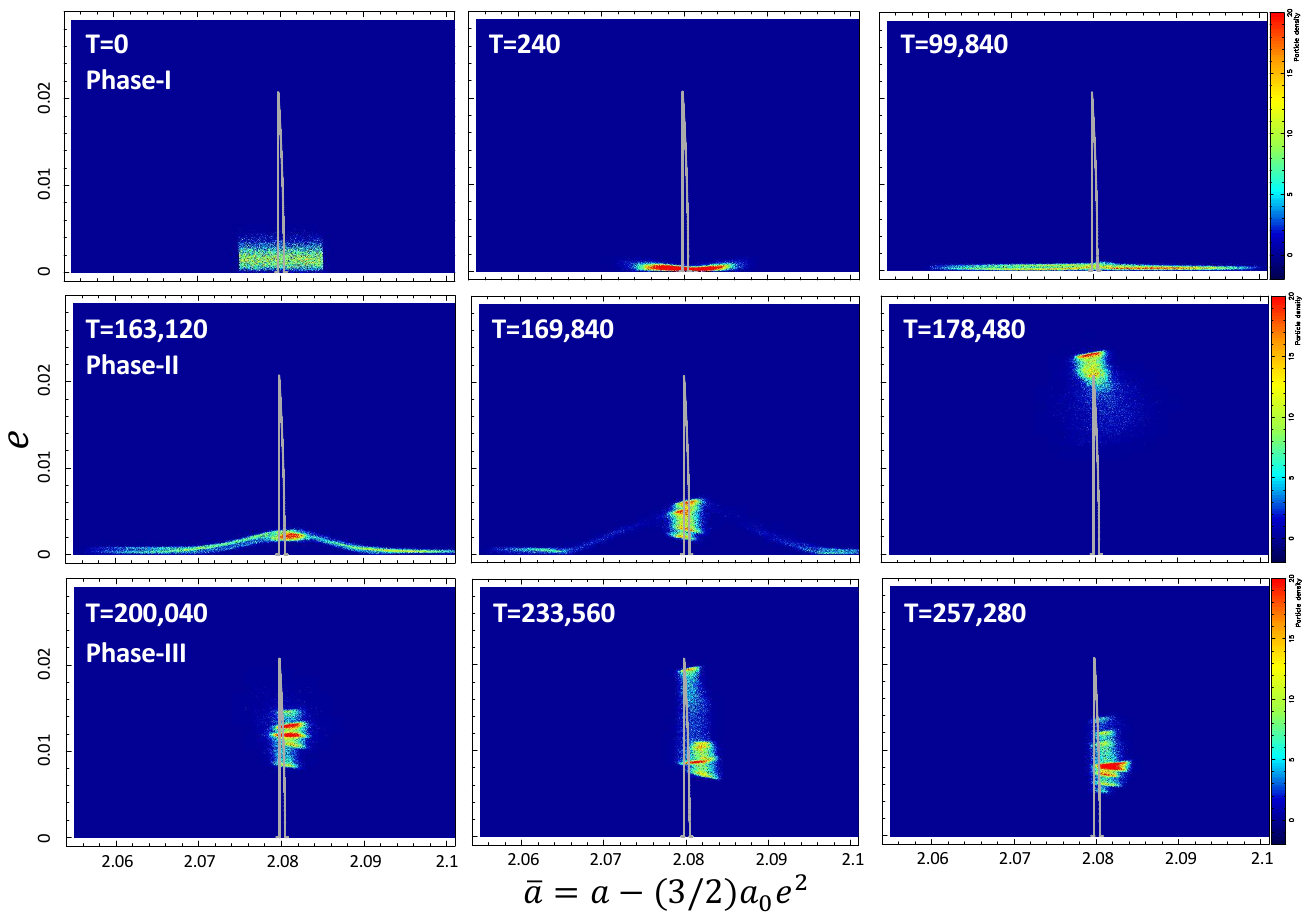}}
\centerline{\includegraphics[width=200mm]{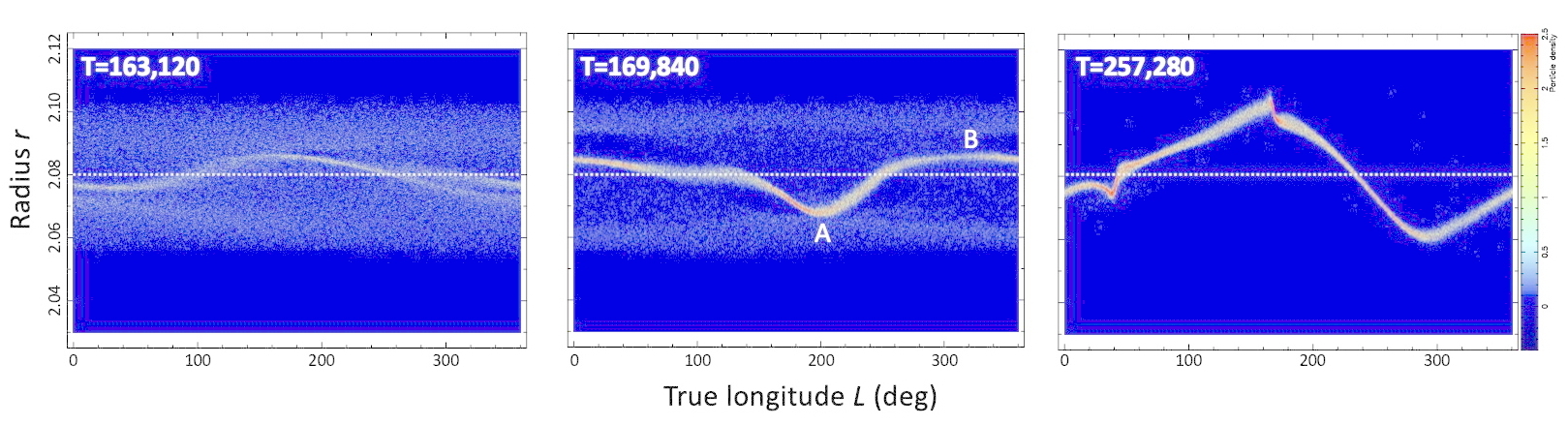}}
\caption{
A simulation with 40,000 ring particles of radius $R=1.25 \times 10^{-4}$ (corresponding to 25~m for Chariklo's ring particles) perturbed by a mass anomaly $\mu=3 \times 10^{-3}$. The time $T$ is given in units of Chariklo's rotation period (about 7~h, Table~\ref{tab_param_cha}), so that the maximum time shown here ($T=$~256,280) corresponds to about 206~years. \textit{Upper nine panels}:
the density maps of the particles in the modified semimajor-eccentricity $(\overline{a},e)$ space, see Eq.~\ref{eq_modified_a}.
The gray spiky curve is the value of $e_{\rm max}$ shown in the lower left panel of Fig.~\ref{fig_num_vs_theo_scalings}.
Three phases of the ring evolution are displayed. The Phase-I corresponds to a rapid damping of the initial eccentricities accompanied by a radial spreading caused by collisions; During Phase-II, the 1/3 SOR excites the orbital eccentricities of the particles up to the predicted peak value $e_{\rm peak}$ (Eq.~\ref{eq_peak}), while confining concomitantly the material near the resonance ($\overline{a} \approx 2.08$) over a timescale $T_{\rm res}$ (Eq.~\ref{eq_tres}). Finally, during Phase-III, the eccentricities damp due to dissipative collisions and a quasi steady-state is reached, where the  eccentricities damping is balanced by the resonance excitation.
\textit{Lower three panels}:
the density maps of the particles in polar coordinates $(L,r)$ space at three selected times shown in the upper panels. The white dotted lines indicate the location of the 1/3 SOR.
At $T=$~163,120 a streamline forced by the 1/3 SOR has appeared, with a tendency of self-crossing around $L=260$~deg.
It is gathering material from the background unexcited particles.
At $T=$~169,840, the accumulation process is going on, where a well-formed ringlet with various azimuthal modes collects more background material.
At $T=$~257,280, all the ring material as been accumulated onto the ringlet, which is now dominated by a $m=1$ azimuthal mode, with the presence of two kinks.
 The colors in the plots indicate the density of particles (in arbitrary units, density increasing from blue to red). 
Movies generated from these snapshots are available online.
}
\label{fig_map_J_e_snapshots_phases_I_II_III}
\end{figure*}

\subsection{The ring confinement at the 1/3 resonance}

We now describe the various phases of the ring confinement process, using as an example a simulation with more realistic parameter values. The particles have a radius $R=1.25 \times 10^{-4}$ (about 25~m for Chariklo's ring particles) and are perturbed by a mass anomaly $\mu=3 \times 10^{-3}$.
Due to smaller $\mu$ the timescale of the evolution is much longer than in the simulation of Fig.~\ref{fig_combined_effect_collision_sor}. For example, it takes about 150,000 central body revolutions before the resonance accumulation starts. The total span of the simulation, $T \approx$~257,000 corresponds to about 206 years in case of Chariklo.

\begin{figure}[!t]
\centerline{\includegraphics[width=\columnwidth]{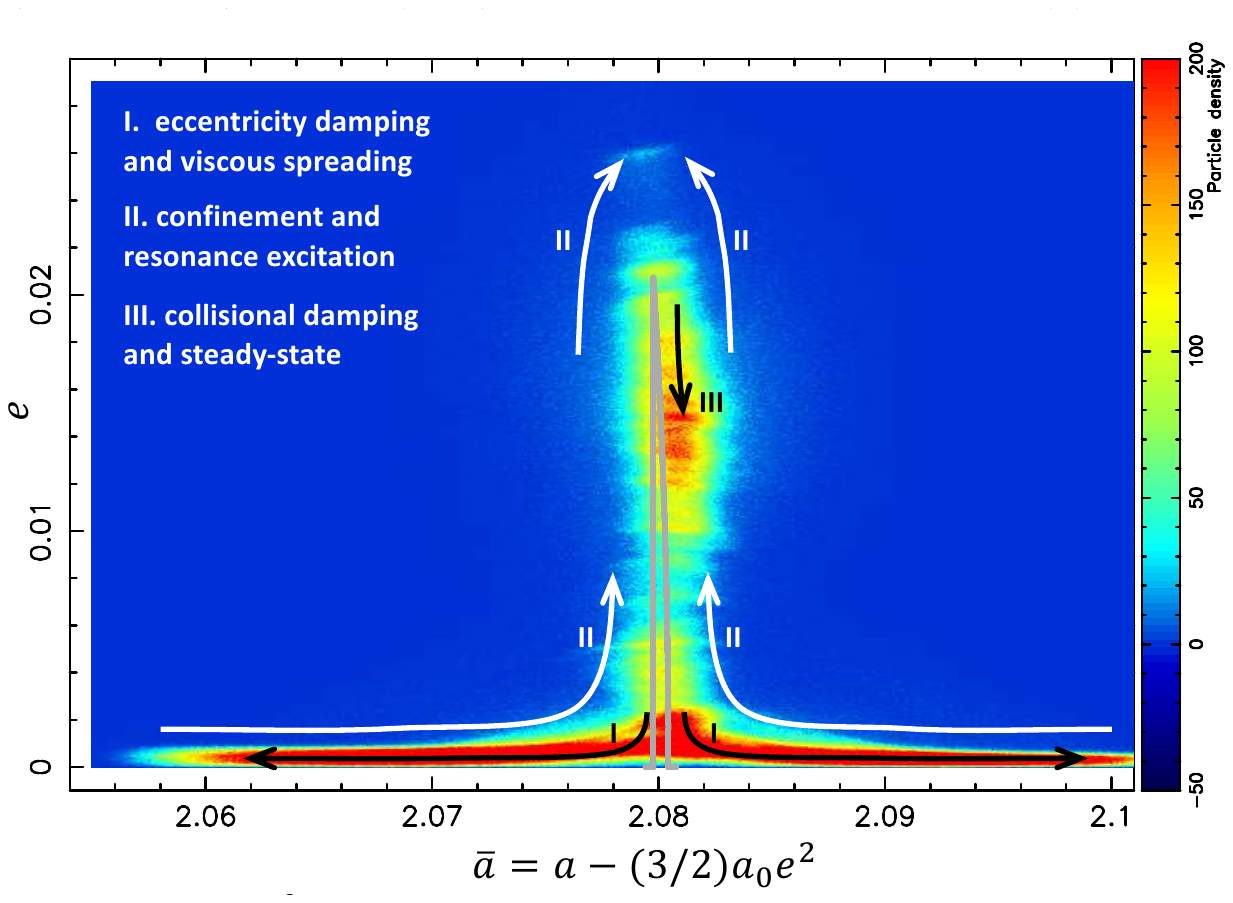}}
\centerline{\includegraphics[width=\columnwidth]{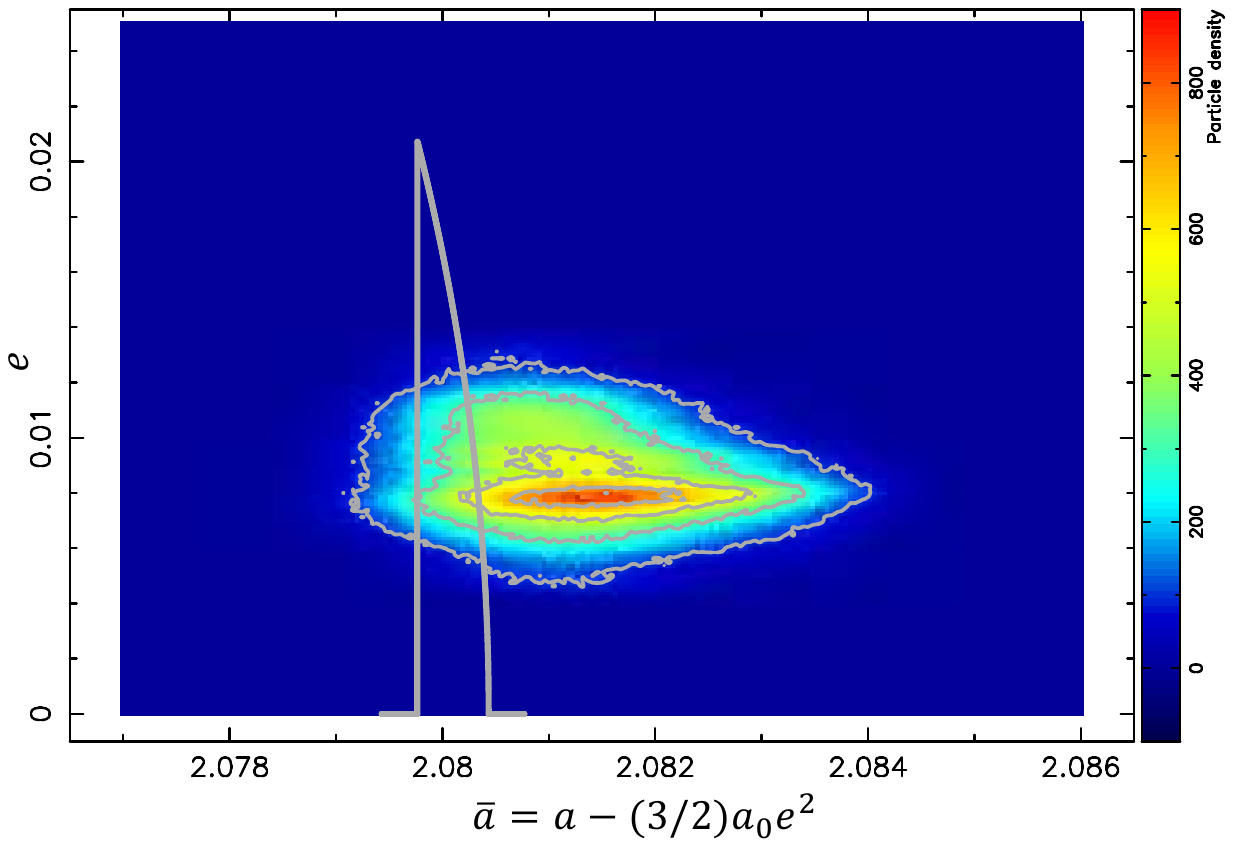}}
\caption{
\textit{Upper panel}:
An overview of the three phases shown in the upper panels of Fig.~\ref{fig_map_J_e_snapshots_phases_I_II_III}.
The plot shows a stack of 1,600 snapshots of the system from $T=$~120,000 to $T=$~200,000 (96 to 160~years respectively in the Chariklo case) with time steps $\Delta T=50$ Chariklo's revolutions (about 15~days). The three phases I, II and III described in the text are indicated by the arrows.
\textit{Lower panel}:
The phase III quasi-stationary stage obtained at the end of our run (note the change of radial scale compared with the upper panel). A total of 780 snapshots taken from $T=$~256,500 to $T=$~257,280 (204.94 to 205.57 years) with time steps $\Delta T=1$ Chariklo's revolution (about 7~h) have been stacked. A quasi steady state is reached, where the eccentricity damping due to collisions is balanced by the resonance excitation. On the long term (nor reached here) a leakage of particles towards larger semimajor axes would be observed, as illustrated in Fig.~\ref{fig_effect_satellite}, where a larger mass anomaly $\mu=0.03$ is used.
}
\label{fig_map_J_e_stacked}
\end{figure}

The nine upper panels of Fig.~\ref{fig_map_J_e_snapshots_phases_I_II_III} display three phases of the ring evolution.
Phase-I corresponds to the damping of the initial eccentricities followed by a radial viscous spreading of the ring material; 
during Phase-II, the 1/3 resonance excites the orbital eccentricities of the particles,
forming a ringlet which gathers material from the background material. The eccentricity then reaches the peak value $e_{\rm peak}$ (Eq.~\ref{eq_peak}). A similar surge of eccentricities and its ensuing damping were visible in a simulation with $\mu=0.1$, see the red line in the lower left panel of Fig.~\ref{fig_combined_effect_collision_sor}.
This damping leads to a quasi steady-state that constitutes the Phase-III. A ringlet is now confined, in which the damping of eccentricities by collisions is balanced by the resonance excitation.
The Fig.~\ref{fig_map_J_e_stacked} synthesizes the three phases, in which 1600 snapshots of the system have been stacked in the $(\overline{a},e)$-space. It summarizes the history of the ring material, from the initial viscous spreading phase to the steady-state situation.

The three lower panels of Fig.~\ref{fig_map_J_e_snapshots_phases_I_II_III} show the ring in polar coordinates at three selected times. 
At $T=$~163,120, a streamline excited by the 1/3 SOR has appeared, exhibiting a tendency of self-crossing as expected from  the Fig.~6 of  Paper~I. At $T=$~169,840, this streamline is gathering particles from the unexcited background ring material.
This gathering can be understood by the fact that the background particles near the point $A$ move slower than the particles in the streamline. Consequently, they receive angular momentum during collisions, so that their semimajor axes increase. Conversely, background particles near the point $B$ see their semimajor axes decrease during collisions. This globally leads to the confinement of the particles near the resonance radius. At $T=$~257,280, all the background particles have been gathered into a non-intersecting streamline dominated by a $m=1$ azimuthal mode.

The results of a complementary run using a larger mass anomaly $\mu=0.1$ are provided in Fig.~\ref{fig_wide_initial_distribution}. It shows better the initial self-intersecting streamline forced by the 1/3 SOR, and its further transformation into a non-intersecting streamline dominated by a $m=1$ azimuthal mode. In this case, two other ringlets form inside and outside the central ringlet, the outer ringlet being eventually swallowed by the central ringlet during the run.

\subsection{Scaling to physical systems}
\label{sec_scaling} 
We now address the applicability of simulation results to real systems. Firstly, we study the role of impact frequency in the transition between a test-particle system and a collisional ring. Then, we provide the criterion for which the resonant confinement is expected to win over the viscous spreading due to collisions. These are important factors since a fully realistic simulation of an azimuthally complete dense ring including collisions between presumably meter-sized particles would imply an unmanageable number of such particles\footnote{For example, Chariklo's CR1 ring, with radius of
  about 400 km and mean width 7.5 km, has a surface area of $\sim 2  \times 10^{10}$~m$^2$, corresponding to a total area equivalent to
  $\sim 6 \times 10^9 $ one-meter radius particles.}.  In practice, a smaller optical depth and larger than real particles must be employed in simulations, calling for a scaling between the particle size $R$, mass anomaly $\mu$, and optical depth $\tau$ in simulated and real physical systems.

In the non-gravitating simulations with constant rebound coefficient $\epsilon_{\rm n}$, the collisional steady-state of a non-perturbed simulation system is determined by three parameters: the particle radius $R$, the dynamical optical depth $\tau$, and the coefficient of restitution $\epsilon_{\rm n}$. 
The influence of perturbation depends on the mass anomaly $\mu$ and on the particular resonance(s) acting on the ring.  
For a non-axisymmetric central body, the strength of the perturbation also depends on the oblateness $f$ and elongation $\epsilon_{\rm elon}$.  
To a lesser extent, the initial radial width of ring $W_0$ also matters, as it determines the maximum number of particles that can be perturbed by the resonance. Ideally, $W_0$ should be large compared to both $R$ and the resonance width $W_{\rm res}$. 
On the other hand, the initial values of the vertical ring thickness and velocity dispersion are not critical, since the  collisional damping of eccentricities and inclinations rapidly leads in the non-perturbed case to a steady-state with velocity dispersion $c \sim R n$, the precise proportionality factor depending on $\epsilon_{\rm n}$ \citep{salo2018}.  The importance of $\tau$ follows from the fact that in a non-perturbed $3D$ ring the impact frequency $w_{\rm c}$ is proportional to $\tau n$. The basic formula for viscosity is $\nu \approx w_{\rm c} \lambda^2$, where the radial mean free path $\lambda \approx c/n$ at low $\tau$ rings. This implies  
\begin{equation}
    \nu = k_{\rm visc} \tau R^2 n,
    \label{eq:kvisc}
\end{equation}
\noindent where $k_{\rm visc}$ is a numerical factor of the order of unity. With our standard value $\epsilon_{\rm n}=0.1$, we have 
$k_{\rm visc} \approx 3.5$. 

When scaling the simulations to physical systems, two questions  at  least arise:
(1) What is the minimum frequency of impacts $w_{\rm c}$ that  makes the ring behave as a collisional ring, in contrast to a mere ensemble of independent test-particles?
(2) What is the parameter regime in which the timescale for the resonance build-up is shorter than the viscous spreading time due to collisions?
These two questions are addressed in the next two subsections.

\begin{figure*}[!ht]
\includegraphics[width=2\columnwidth]{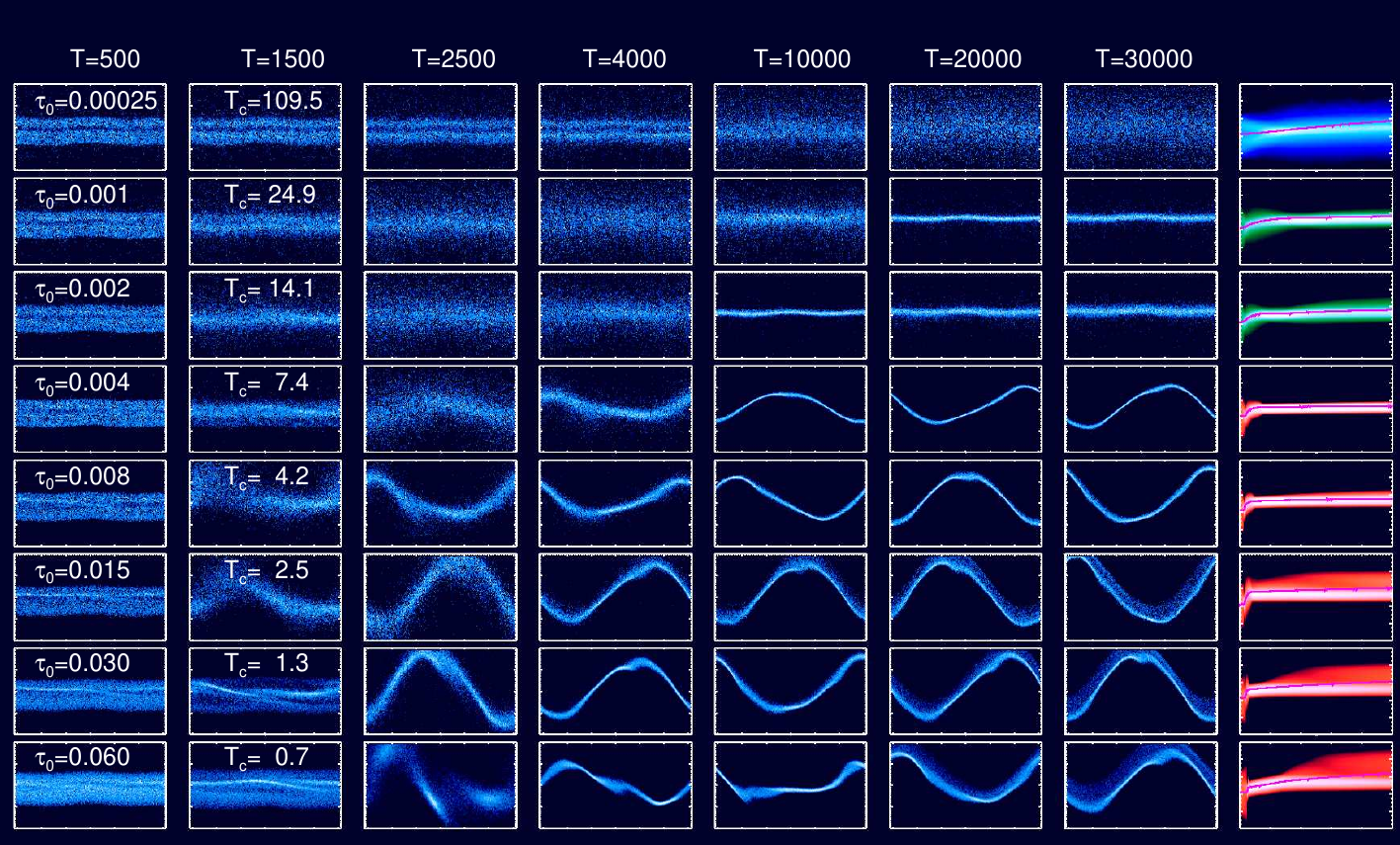}
 \caption{
 The transition from non-collisional to collisional rings. 
 This is explored by simulations with various initial optical depths $\tau_0$ from $0.00025$ to $0.06$, all using a mass anomaly $\mu=0.1$. Snapshots of the particle distributions in polar coordinates are shown at various times, using the radial range 1.88-2.28 centered around the resonant semimajor axis $a_{1/3} \approx 2.08$. The label $T_{\rm c}$ in the second column indicates the average time interval between impacts, in units of the particle orbital periods. The rightmost column shows the time evolution of the vertical angular momentum $L_z$ distribution with time up to 30,000 central body revolutions. The lowermost simulation ($\tau_0=0.06$) is the same as the one shown in the last row of Fig.~\ref{fig_combined_effect_collision_sor}, where it is displayed up to $T=5000$.
 }
\label{fig_transition_collisionless_collision}
\end{figure*}

\subsubsection{Influence of impact frequency}

For non-gravitating rings, the impact frequency $w_{\rm c} \approx 3 n \tau$, corresponding to about $20 \tau$ impacts/particle/ring orbital period. A condition sometimes quoted for a ``collisional ring" is to have at least one impact/particle/ring orbital period. This condition, $w_{\rm c} > n/(2\pi)$, would correspond to $\tau \gtrsim 0.05$. However, since the resonance timescales at 1/3 SOR are much longer than the ring orbital period (see Eq.~\ref{eq_tres}), a much smaller impact frequency can in practice be expected to modify the test particle evolution. 

We explore the transition from a non-collisional to a collisional ring by conducting simulations with successively larger values of initial optical depths $\tau_0$. Fig.~\ref{fig_transition_collisionless_collision} shows snapshots of the ring near the 1/3 SOR, using $\mu=0.1$ and $\tau_0 =0.00025 - 0.06$. The corresponding unperturbed impact frequency $w_{\rm c}$ increases from about 0.01 to about one impact/particle/orbit, indicating on average one-hundred to one ring orbital periods between impacts (marked as $T_{\rm c}$ in the second column). For the second-order 1/3 SOR, $T_{\rm res} \sim 400$ central body rotation periods (Eq.~\ref{eq_tres}), i.e. about 130 ring orbital periods. This timescale refers to the time it takes to reach the maximum eccentricity starting from circular orbits; the e-folding time $T_{\rm exp}$ of eccentricity growth is about ten times shorter.

Thus, for the smallest $\tau_0$ explored in Fig.~\ref{fig_transition_collisionless_collision}, the impact timescale $T_{\rm c}$ is roughly equal to the resonance timescale $T_{\rm res}$. Consequently, the effect of impacts is weak: most of the particles initially close to the resonance are able to cross radially the ring without colliding. A gap similar to that in the upper panel of Fig.~\ref{fig_combined_effect_collision_sor} (collisionless test particles) opens at the resonance. This gap gets slowly filled with time, as inter-particle impacts lead to increased eccentricities throughout the ring. No sign of particle accumulation in the resonance region is seen during the time span of the simulation. In contrast, the ring appears very diffuse. 

When $\tau$ increases to $\tau=0.001$ and $\tau=0.002$ (corresponding to the $L_z$ maps colored green in the last column of Fig.~\ref{fig_transition_collisionless_collision}),  both the opening of the initial gap and the rapid growth of eccentricities take place. The system eventually forms a nearly circular ringlet just outside of the $1/3$ resonance, surrounded by a population of ``hot" ring particles not trapped by the resonance. We note a weak undulation of the ringlet with azimuthal number $m=2$, caused by the distant 2/3 resonance which is still quite strong at the 1/3 SOR distance (see Fig.~\ref{fig_resonance_overlap}). 

For $\tau \geq 0.004$, corresponding to $T_{\rm c} \lesssim 10$, the behavior is reminiscent of that of Fig.~\ref{fig_combined_effect_collision_sor}: the whole system goes through the initial accumulation, resonance excitation and confinement phases ($L_z$ maps colored in red in the last column).
The eventual leakage of particles outside the resonance radius is obvious for the $\tau \gtrsim 0.01$ runs, the dispersal of the ringlet getting faster with larger $\tau$.

To conclude, all the simulations of Fig.~\ref{fig_combined_effect_collision_sor} where $T_{\rm c}  \lesssim 0.1 T_{\rm res} \approx T_{\rm exp}$ show an evolution similar to what is displayed in Figs. \ref{fig_map_J_e_snapshots_phases_I_II_III} and \ref{fig_map_J_e_stacked}. Taking into account the scaling between $T_{\rm res}$ and $\mu$ leads to the following empirical condition for a ring near the 1/3 SOR to be collisional (instead of an ensemble of test particles),

\begin{equation}
    \tau \gtrsim (0.01-0.04) \mu.
    \label{eq:taumin}
\end{equation}
Here the larger limit corresponds to the formation of confined eccentric ringlet while the lower limit corresponds to $T_c$ where the  first signs of collective behavior appear.

\subsubsection{Competition between resonance accumulation and viscous spreading}

\begin{figure*}[h]
  \includegraphics[width=2\columnwidth]{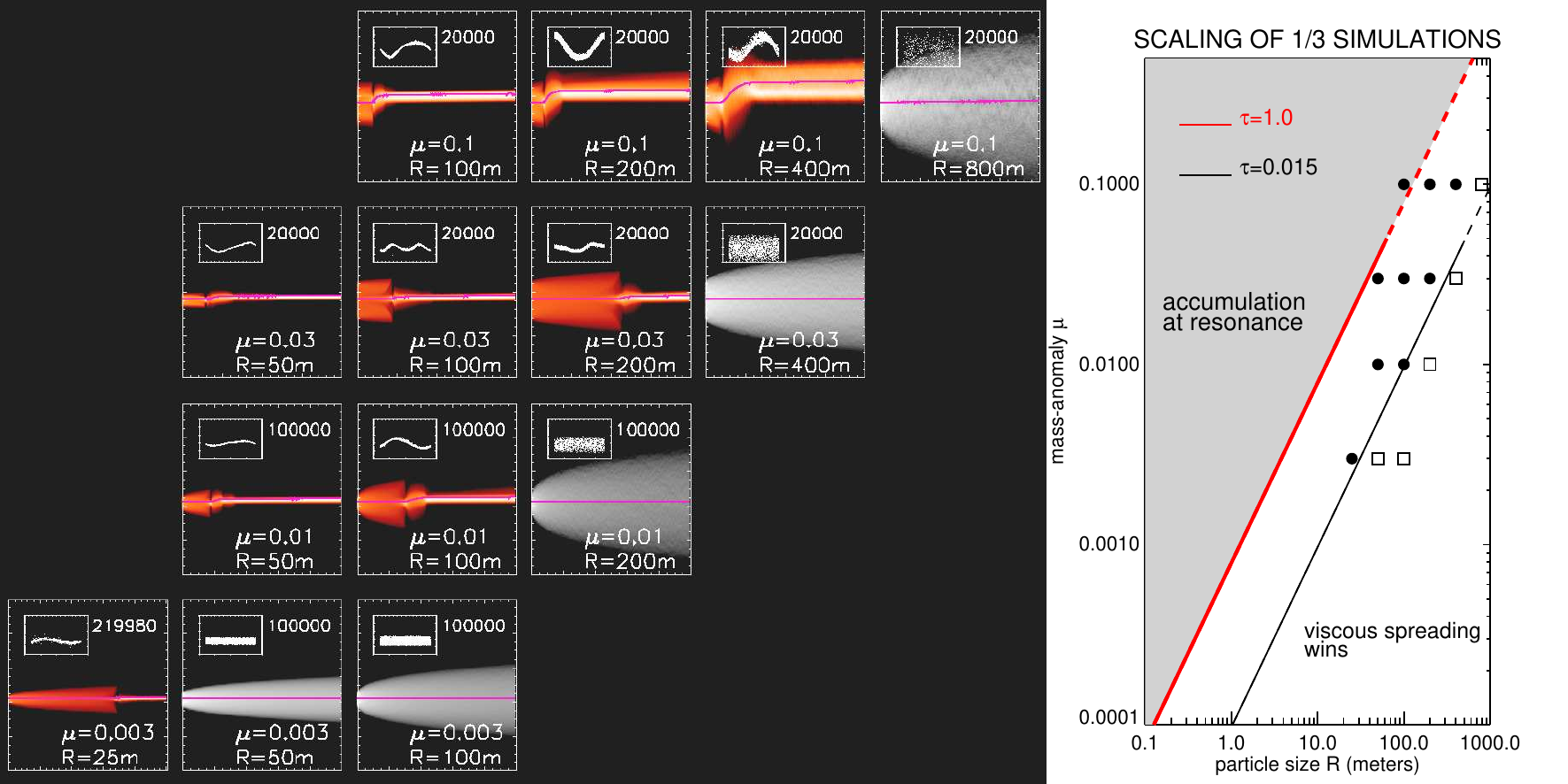}
  \caption{
  The transition between accumulation and dispersal, a test of Eq.~\ref{eq_threshold}. 
  \textit{Left panel}:
  The same as the left column of Fig.~\ref{fig_combined_effect_collision_sor} exploring a grid of simulations near the 1/3 SOR. The particle radii increase from left to right in the range $R=1.25 \times 10^{-4}-4 \times 10^{-3}$, corresponding to $R=25-800$~m in the case of Chariklo's rings, while the mass anomaly increases upward from 0.003 to 0.1. The number of particles and the initial width of the rings have been chosen so that to provide the same initial optical depth $\tau_0 = 0.015$ for all simulations. The vertical axes span values of $L_z$  between 1.35 and 1.55 for $\mu=0.1$ and $0.03$, and between 1.4 and 1.5 for $\mu=0.01$ and $0.003$. The small inserts covering radial range 1.9-2.4 show snapshots of the ring in polar coordinate at the end of the run, with the label indicating the duration of the run in Chariklo's rotation periods. Increasing the particle size (and thus the viscosity) for a given perturbation strength prevents the resonance accumulation. 
  \textit{Right panel}:
  The filled/open symbols distinguish between simulations leading to resonance confinement/dispersal, respectively. The black line indicates the accumulation threshold for the simulations displayed in the left panel, following Eq.~\ref{eq_threshold}; the dashed portion of the line indicates the region where the linear relation fails since the resonance excitation at large $\mu \gtrsim 0.03$ becomes somewhat weaker than predicted by the analytical formulas (see footnote~1).    The shaded region bounded by the red line extrapolates the accumulation region to a denser ring with $\tau=1$.
  }
  \label{fig_threshold}
\end{figure*}

The eccentricity growth forced by a resonance takes place on timescales $\sim T_{\rm res}$. The growing epicyclic excursions induce collisions between particles in and close to the resonance zone. Due to the dissipative nature of the impacts, the particles originating further away from the resonance tend to remain closer to resonance after impacts as their eccentricities are damped. In order to accumulate particles at the resonance, the viscous removal of particles from resonance must not be too fast. To get a condition for net accumulation, we follow the heuristic arguments presented in \cite{franklin1980}.
    
The displacements of particles diffusing away from resonance zone are given by
\begin{equation}
 \Delta r_{\rm visc} \approx \sqrt{\nu t},
\end{equation}
while the epicyclic excursions due to growing eccentricities transport particles to the resonance zone from distances
\begin{equation}
\Delta r_{\rm res} = e_{\rm res}(t) a_0. 
\end{equation}
These particles preferentially collide with particles near the resonance zone and end up with mean distances close to the resonance. In order to obtain a net accumulation at the resonance, we must have $\Delta r_{\rm visc} \lesssim \Delta r_{\rm res}$ during the whole time range $t \lesssim T_{\rm res}$. This implies $\sqrt{\nu T_{\rm res}} \lesssim e_{\rm max} a_0$, leading to the condition
\begin{equation}
    \nu \lesssim (e_{\rm max} a_0)^2/T_{\rm res}= k_{\rm res} (e_{\rm peak} a_0)^2/{T_{\rm res}}.
\end{equation}
Here, the prefactor $k_{\rm res} << 1$ takes into account the fact that the average $e_{\rm max}$ amplitudes near the resonance are less than the peak value $e_{\rm peak}$, and also that the timescales to reach $e_{\rm max}$ are generally longer than $T_{\rm res}$, see the right panel of Fig.~\ref{fig_num_vs_theo_scalings}.
  
We now plug in Eq.~\ref{eq:kvisc} for the viscosity, while using the formulae~\ref{eq_peak} and \ref{eq_tres} for $e_{\rm peak}$ and $T_{\rm res}$ at the 1/3 SOR. This provides a simple condition for the parameter regime in which  the  resonance confinement should be possible,
\begin{equation}
    k {\mu}^2 \gtrsim \tau R^2,
    \label{eq_threshold}
\end{equation}
where $k\sim 0.05 k_{\rm res}/k_{\rm visc}$.  Note that this estimate suggests that the threshold $\mu$ for confinement is expected to scale as $\mu_{\rm thresh} \propto \sqrt{\nu}$. In order to check the $\tau, R$ and $\mu$ dependence of the accumulation threshold and to estimate the value of $k$, we turn to numerical simulations. 

\subsubsection{Empirical criterion for ringlet formation}

Figure~\ref{fig_threshold} displays a grid of 1/3 resonance simulations all with the same initial optical depth $\tau_0=0.015$, using values of $\mu$ between 0.003 and 0.1, and particle radii from $0.000125$ to $0.004$ (25~m to 800~m when scaled to Chariklo's ring system). This figure illustrates the competition between resonance accumulation and viscous dispersal, covering for each $\mu$ a range of runs both leading and not leading to confinement.
In particular, this survey confirms that for a fixed $\tau$, and as predicted by Eq.~\ref{eq_threshold}, the minimum $\mu$ required for the resonance confinement increases linearly with $R$.  Thus, a much smaller size of mass anomaly can be expected to lead to resonance confinement when the particle size and ring viscosity approach more realistic small values. The black unit-slope line in the right panel delineates the boundary between the two regimes. Except for the largest value $\mu=0.1$, the boundary is reasonably well approximated by this unit-slope line, in agreement with Eq.~\ref{eq_threshold}. The boundary provides an estimated value $k \sim 4 \times 10^{-5}$, so that Eq.~\ref{eq_threshold} for 1/3 SOR can be re-expressed as
\begin{equation}
\mu \gtrsim 160 \sqrt{\tau} \ R,
\label{eq_threshold_bis}
\end{equation}
where we recall that the particle radius $R$ is measured in units of the corotation radius $a_{\rm cor}$. For Chariklo system this yields
\begin{equation}
\mu \gtrsim 10^{-3}\sqrt{\tau} \ R_{\rm phys},
\label{eq_threshold_bis_phys}
\end{equation}
where $R_{\rm phys}$ is the particle radius in meters.
As found in test particle integrations, for $\mu=0.1$ the eccentricity amplitudes are about 30\% weaker than predicted by the theoretical formula for $e_{\rm peak}$. It also takes longer to reach the maximum eccentricity than predicted by the fitted $T_{\rm res}$ based on smaller $\mu$'s. Together these effects weaken the perturbation by roughly a factor of two, explaining the reduced threshold value of $R$ for $\mu=0.1$ compared to the black line limit.

Additional tests were performed to check the $\tau$ dependence of the accumulation boundary: according to Eq. \ref{eq_threshold_bis} accumulation requires $\tau \lesssim  \tau_{\rm lim} \approx 4 \times 10^{-5} (\mu/R)^2$.  In case of Fig.~\ref{fig_transition_collisionless_collision} with $\mu=0.1, R=10^{-3}$, the implied $\tau_{\rm lim} \approx 0.4$, consistent with the fact that accumulation was seen even in the case of the largest $\tau_0=0.06$. Appendix  \ref{AppendixC3}  reports similar surveys using $R=10^{-3}$ with $\mu=0.05$ and $0.03$, in which case the predicted $\tau_{\rm lim} \approx 0.1$ and $0.035$, respectively. Simulations covering a range of $\tau_0$'s confirm the expected trend of reduced $\tau_{\rm lim} \propto \mu^2$, though the limiting values observed in simulations are roughly 30\% smaller than estimated, again most likely due to the large $\mu$.

Besides the value $\tau=0.015$ used in the simulations, an extrapolated curve for $\tau=1$ is shown as a red line in the right panel of Fig.~\ref{fig_threshold}. It assumes that the linear relation $\nu \propto \tau$ holds all values of $\tau$, which is not strictly true when $\tau$ starts to approach unity (see \cite{salo2018}). In any case, the red curve indicates that confinement should take place in dense Chariklo type rings provided that mass anomaly exceeds about 0.001, assuming typical 1-meter ring particles. Note that this estimate includes only collisions: see Sect. \ref{sec_sg} for a refined estimate in case gravitational viscosity is taken into account.

\begin{figure}[h]
  \includegraphics[width=1\columnwidth]{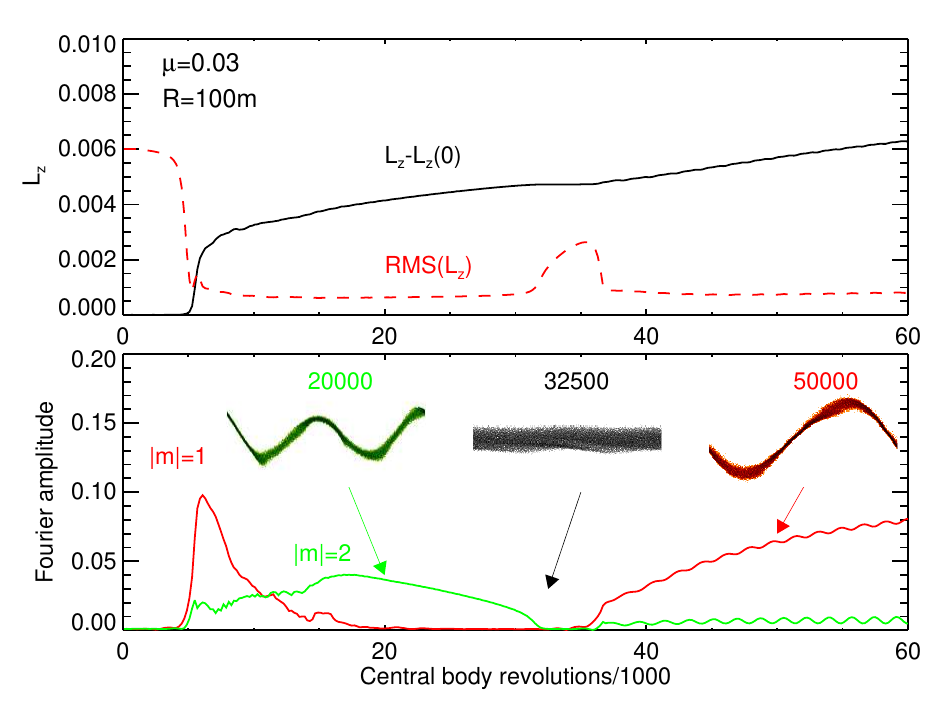}
  \caption{
  The transition from a $|m|=2$ ringlet stage to a $|m|=1$ ringlet stage in the $\mu=0.03$, $R=5\cdot 10^{-4}$ (corresponds to 100~m in case of Chariklo) simulation of Fig.~\ref{fig_threshold}. In all other cases the ringlet shape is dominated by the $|m|=1$ response once it has formed. Also in this case, the $|m|=1$ response eventually replaces the $|m|=2$ shape. During this transition at $T \approx$~35,000, the torque exerted on the nearly circular ringlet temporarily vanishes (see the flat portion of mean $L_z$ curve in the upper frame).
  }
  \label{fig_m1_m2}
  \end{figure}

\section{Normal modes}

In the case of forced Lindblad resonances ($j=1$ in Eq.~\ref{eq_ratio_n_OmegaB}), the response of collisional rings is
relatively straightforward: dissipative impacts force the particles to follow $m$-lobed non-intersecting streamlines that are close to the resonant periodic orbits (See Fig.~\ref{fig_ilr_olr}). In the more general case of a  $m/(m-j)$ SOR, the periodic orbits have $|m|(j-1)$ self-intersecting points \citep{sicardy2020}. Thus, in the case of the 1/3 SOR ($m=-1$, $j=2$), there would be one intersecting point, see  the Fig.~6 of  Paper~I. In practice, collisions prevent this crossing of orbits from persisting (see the $T=3000 - 8000$ frames of Fig.~\ref{fig_wide_initial_distribution}).
An interesting question is what is the resulting flow configuration the system settles to?

According to our simulations, the typical overall outcome is an excitation of a dominant $|m|=1$ mode. For example, among the simulations shown in Fig.~\ref{fig_threshold}, only in one case a $|m|=2$ ringlet was initially formed. Even in this case, when the simulation was continued further, the $|m|=1$ mode eventually took over (Fig.~\ref{fig_m1_m2}). However, besides the $|m|=1$ mode, additional
Fourier modes are always present, see for example the $T=$~30,000 frames in Fig.~\ref{fig_transition_collisionless_collision}. The most striking example is the $\mu=0.003$ simulation of Fig.~\ref{fig_map_J_e_snapshots_phases_I_II_III} which settles to a complicated azimuthal profile with two prominent kinks. As demonstrated below, such a ring response can be interpreted as the 1/3 SOR excitation being transferred to a superposition of several free outer Lindblad modes. Note that in this simulation the smallest particle size and perturbation are used, whereupon it can be assumed to mimic closest the possible behavior of real rings. In what follows we analyze in more detail the ringlet formed in this simulation.

\begin{figure}
  \includegraphics[width=0.9\columnwidth]{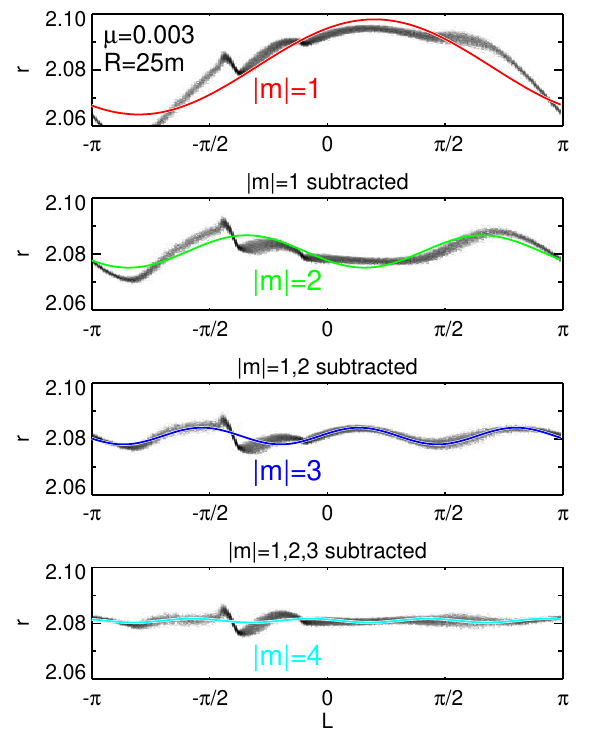}
  \caption{
  Fourier fits to the ring azimuthal profile in the $\mu=0.003$, $R=25$~m simulation at $T=$~250,000 (same simulation as in Fig.~ \ref{fig_map_J_e_snapshots_phases_I_II_III}). The uppermost frame shows $r(L)$ profile, with a $|m|=1$ fit superposed. The second frame shows the residual profile after subtracting the $|m|=1$ component; the green curve is the $|m|=2$ fit to the residual profile. The next frames repeat the procedure for $|m|=3$ and $|m|=4$.
  }
  \label{fig_m1234_subtract}
\end{figure}

\begin{figure}
  \includegraphics[width=1\columnwidth]{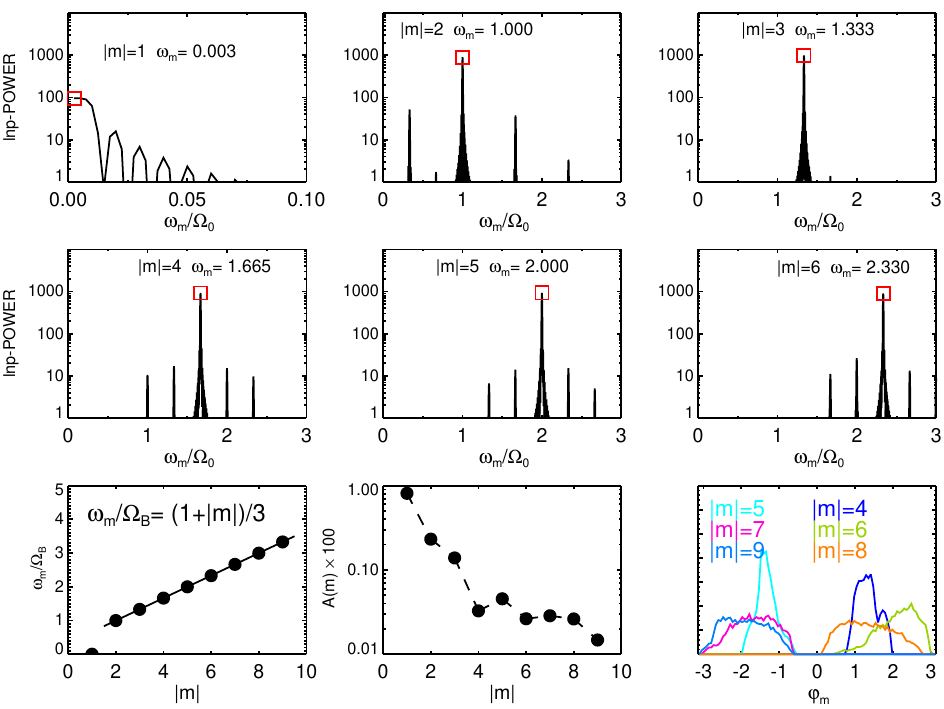}
  \caption{
  The proper Lindblad modes appearing in a ring confined at the 1/3 SOR.
  The outputs of the run shown in Fig.~\ref{fig_map_J_e_snapshots_phases_I_II_III} have been  used to detect normal modes in the confined rings between times 250,000 and 250,200, with steps $\Delta T=0.05$. The Lomb normalized periodogram power spectra of the modes described by Eq.~\ref{eq_model_m_modes} are displayed in the upper six panels as a function of $\omega_m/\Omega_{\rm B}$ for $|m|=1$~to 6. Consistent with Eq.~\ref{eq_possible_proper_modes_2}, the maximum power (red squares) is reached at the frequency $\omega_m = (1+|m|)\Omega_{\rm B}/3$. This is illustrated in the lower left panel.  The lower middle panel shows the rapid decrease of the amplitude $A_m$ of the modes with $|m|$, while the right panel shows the distribution of phases $\theta_m$.
  }
  \label{fig_mode_analysis}
\end{figure}

Figure~\ref{fig_m1234_subtract} shows the radius versus longitude profile of the $\mu=0.003$ simulation, at $T=$~250,000. In the uppermost frame the fitted $|m|=1$ Fourier-component is superposed on the profile, while the lower frames show the residual profile after subtracting consecutive components $|m|=1,2,3$. Clearly, both $|m|=2$ and $|m|=3$ modes are significant besides the $|m|=1$ mode, whereas subtracting the $|m|=4$ mode would not have much effect on the residual profile. In order to analyze the propagation of the modes, a shape model
\begin{equation}
r_m(L,t) = A_m \cos [m(L - \Omega_m t) + \phi_m], 
\label{eq_model_m_modes}
\end{equation}
was fitted to each Fourier component over time range $\Delta T=200$. Here  $r_m(L,t)$ is the radius of the streamline versus the true longitude $L$ at time $t$, $A_m$ is the amplitude of the mode and $\phi_m$ its phase, $\Omega_m$ is the pattern speed and $\omega_m= |m|\Omega_m$ is the frequency. The results of this fit are collected to Fig.~\ref{fig_mode_analysis}.
The peaks of the periodogram for $|m| =2,3,...$ all fall on a linear relation $\omega_m/\Omega_{\rm B} = (1+|m|)/3$, while for $|m|=1$, $\omega_m \approx 0$. 

This is a quite remarkable result as it shows that the system's response is a superposition of free outer (since $\Omega_m > n$) Lindblad modes corresponding to the condition $\kappa= m(n-\Omega_m)$, with $m < -1$. In such a mode, a particle executes exactly $|m|$ radial excursions while completing one revolution in a frame rotating with the pattern speed $\Omega_m$. The $m=-1$ mode corresponds to $\Omega_m= \dot{\varpi}$, i.e. the locked precession mode of an essentially Keplerian ellipse. The other modes with $m \neq 1$ correspond to 
\begin{eqnarray}
 \frac{\Omega_m}{n} \approx \frac{m-1}{m}, \label{eq_possible_proper_modes_1} \\
 \nonumber \\
 \omega_m \approx |m-1|n, \label{eq_possible_proper_modes_2}
\end{eqnarray}
where the approximation stems from the fact that $\dot{\varpi} \ll n$. Taking into account that $n= \Omega_B/3$ and $m < 0$, this is exactly the same relation as observed in the simulation.

Figure~\ref{fig_mode_analysis} indicates that the Fourier-modes with $|m| \ge 4$ have small amplitudes, compared to $|m|=1,2,3$. However, their amplitudes decay quite slowly,  and most importantly, do not stem from noise but convey a true signal, corresponding to the ``kink", best seen in the lowermost residual profile of Fig.~\ref{fig_m1234_subtract}. 

\begin{figure}
\centering
  \includegraphics[width=1\columnwidth]{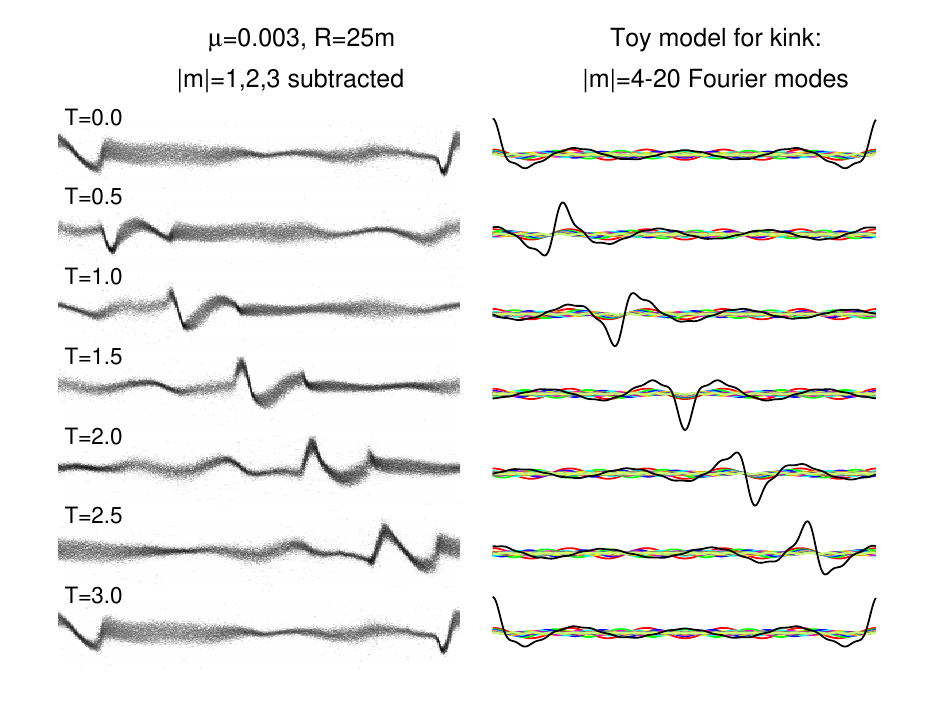}
  \caption{
  \textit{Left column}: the time evolution of the ringlet in the end of the $\mu=0.003$ simulation. The dominant $|m|=1,2,3$ modes have been subtracted in order to highlight the kink feature. 
  \textit{Right column}: a toy model for the formation of a kink as a superposition of $|m|=$~4-20 modes. Each mode has the same phase $\theta_m$  while the amplitudes obey $A_m \propto 1/|m|^{3/2}$. Profiles calculated with Eq. \ref{eq_model_m_modes} are shown over one full ring period.
  }
  \label{fig_toy_model}
\end{figure}

This correspondence is illustrated in Fig.~\ref{fig_toy_model}, comparing the simulated ringlet to a toy model where $|m|=$~4-20 Fourier modes have been superposed. We assume here that the amplitudes drop as $A_m \sim |m|^{-3/2}$ and that the phases are the same for all modes, roughly corresponding to the properties of the simulated modes (the $\pi$ phase shift between $\theta_m$ of even and odd modes seen in last frame of \ref{fig_mode_analysis} only moves the phase of the kink feature, not affecting its shape or propagation; on the other hand, a non-alignment of $\theta_m$'s would destroy the kink feature). The result is a propagating kink, moving with the orbital speed of the ring, closely resembling what is seen in the simulation.
 
\begin{figure}
\centering
\includegraphics[width=1.\columnwidth]{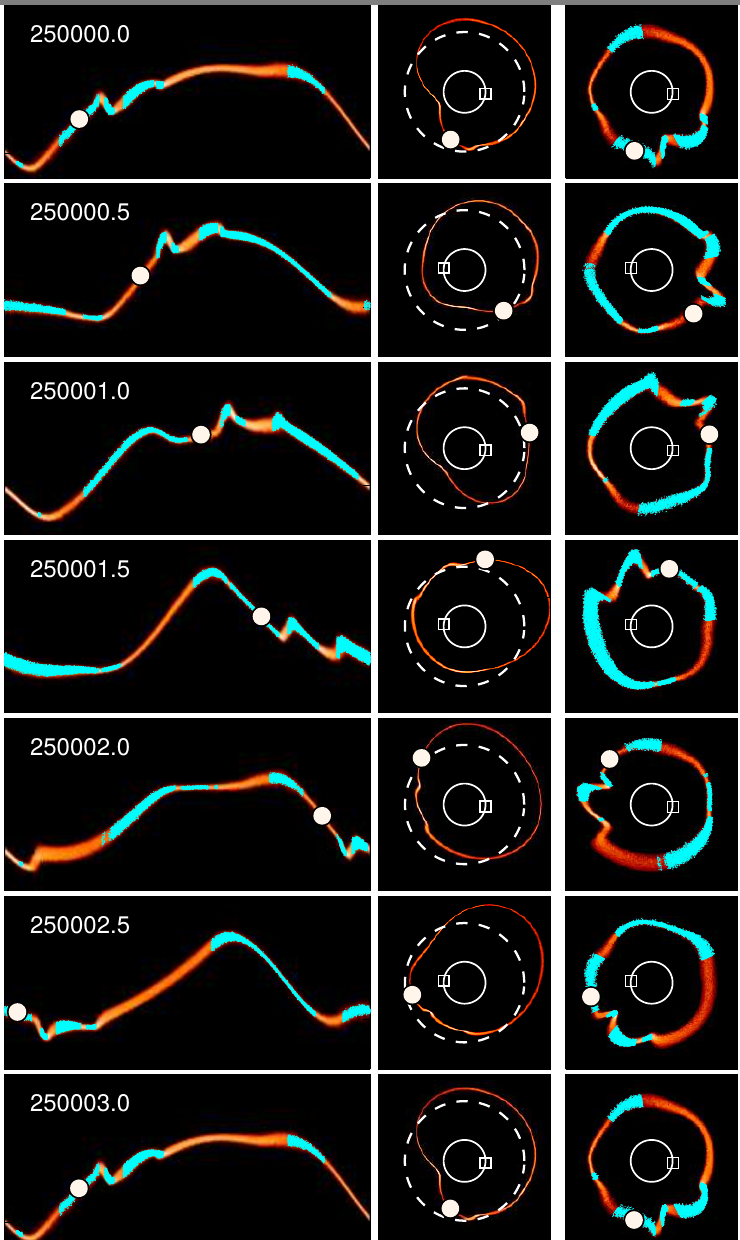}
  \caption{
  Local shear reversal in 1/3 SOR ringlet. The evolution of the $\mu=0.003, R=25$~m simulation is shown over one ring orbital period (or 3 central body revolutions).  The left frames display the $r(L)$ profiles, with radial range 2.05-2.12. The ring patches with negative angular momentum flux $T_{rt}$ are indicated with light blue color, while the white bullet indicates the location of a tracer particle. 
  The middle frames show cartesian projections, where the ring deviations from its mean distance (dashed curve) have been exaggerated by a factor of 40 at each azimuth and the open square marks the location of the mass anomaly.
  Similarly, in the right frames (again a cartesian projection) the ring width variations have been exaggerated by a factor of 50, using the same color convention as in the left frames. For constructing all the figures, twenty snapshots of the ring separated by $\Delta T=3,6,...57$ have been stacked.
  }
  \label{fig_confinement_mechanism}
\end{figure}

Since $\omega_m$ is a multiple of $n$, each mode is invariant through a time translation of $T_{\rm orb}$, the orbital period of the particle. In the present case (a ring near the 1/3 SOR), this means that the ring recovers its initial shape after three rotations of the central body.  We have made use of this in Fig. \ref{fig_confinement_mechanism}, illustrating the ring evolution over one full ring period (three central body rotations), using for each time stacks of 20 particle snapshots separated by $\Delta T=3$. Besides azimuthal profiles, two sets of cartesian projections are shown: in the middle frames the deviations of the ringlet mean distance are exaggerated, while in the right the same is done with width variations. As seen, the shape and width of the ringlet varies in a complicated cycle, however repeating regularly over time.

Also indicated in Fig.~\ref{fig_confinement_mechanism}, are the regions where local shear reversal is taking place, defined as the locations where the non-diagonal component of the velocity dispersion tensor, $T_{rt}=<\Delta v_r \Delta v_t>$, has negative values. Here $\Delta v_r$ and $\Delta v_t$ are the particle velocities with respect to the mean flow at their position, and brackets indicate a mean over a local region. In order to measure  $T_{rt}$ we used a method quite similar to that in \cite{hanninen1992}. We divided the particles into twenty ``streamlines" according to their Jacobi energies, and calculated the mean radial and tangential flow velocities along each such streamline, to finally obtain the $T_{rt}$ along the streamline. Stacking of several snapshots was essential to reduce the noise in this process. 

In a non-perturbed case with Keplerian shear, $T_{rt}>0$ throughout the ring, indicating that collisions on average transfer angular momentum outward (for example, impacts by inner particles approaching with positive relative $v_r$ have on average positive $v_t$ with respect to outer particle, providing a positive impulse, etc.). This situation corresponds to viscous spreading of non-perturbed rings.
Negative values on the other hand associate with inward transport \citep{borderies1983}. The flux of angular momentum (per unit length of streamline) relates to $T_{rt}$ by  $F(a, \phi) = \Sigma(a, \phi) aT_{rt}(a,\phi)$, and the flux integrated over the streamline gives the angular momentum luminosity $L_H(a)$. The situation where $L_H(a)<0$ corresponds to the maintenance of sharp edges due flux reversal \citep{borderies1983} or ``single-sided shepherding" \citep{goldreich1995}. The numerical simulations of \cite{hanninen1994} confirmed that such a mechanism can maintain sharp edges at first-order Lindblad resonances.

Compared to the Lindblad-resonance case  (see Fig.~\ref {fig_ilr_olr}), where the flow pattern, and the regions of positive and negative $T_{rt}$ are fixed in a frame rotating with the satellite, the current situation is more complicated, due to superposition of several modes making the pattern truly variable in time. Because of this the detailed analysis of the angular momentum transport is left for a later study.

\section{Migration of material from inner regions}

So far, we have focused our attention to the confinement of ring material initially near the 1/3 SOR. However, the various rings observed so far around small objects were probably formed over a broad range of radii around the central body, and in very different contexts \citep{sicardy2025a}. For instance, Haumea's ring may have formed during the spewing of material associated with a spin up phase \citep{noviello2022}, while Chariklo's rings may originate from a cometary activity that launched material from its surface \citep{sicardy2025a}.

In general, the resonances rapidly clear the corotation region, pulling the ring material inside the synchronous orbit down to the surface of the body, while pushing away the material outside the synchronous orbit. Using a toy model with Stokes-like friction, \cite{sicardy2019} showed that this clearing may occur over decadal scales under the effect of a Chariklo elongation $\varepsilon_{\rm elon}=0.2$, and in a few centuries if a mass anomaly of $\mu \sim 10^{-3}$ is present. 

\begin{figure}
 \includegraphics[width=1.\columnwidth]{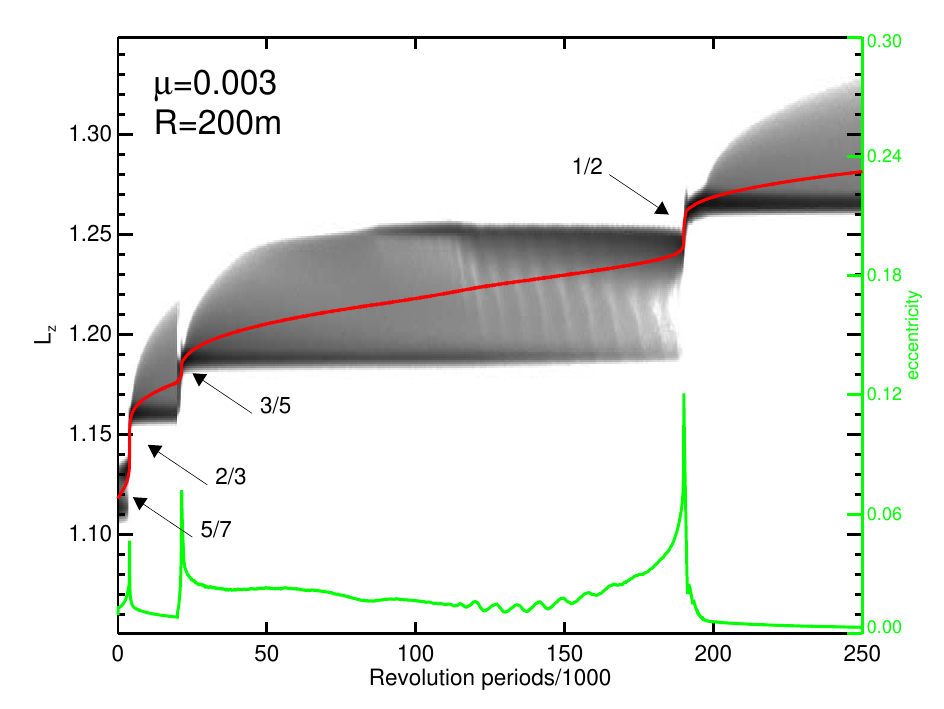}
  \caption{
  The time evolution of the vertical angular momentum $L_z$ of particles that cross various resonances.
  The simulation contains 7,500 particles with radius $R=0.001$ (corresponds to 200~m in case of Chariklo), perturbed by a mass anomaly $\mu=0.003$. The ring starts near the 5/7 SOR and is pushed outwards by a positive torque which takes it through the 2/3, 3/5 and 1/2 SORs. The red curve is the mean value of $L_z$. The green curve shows the mean eccentricity of the particles.
  }
  \label{fig_passage_resonances_t_Lz}
\end{figure}

\begin{figure*}
 \includegraphics[width=2.\columnwidth]{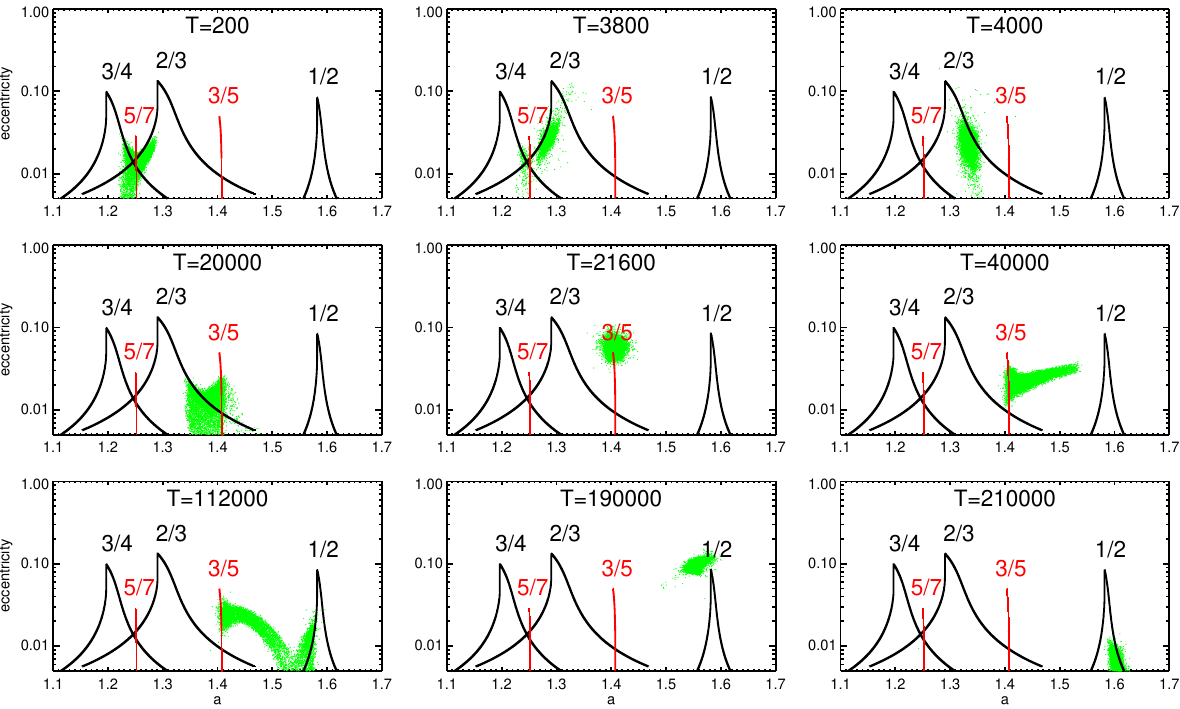}
  \caption{
  The eccentricities $e$ of particles versus semimajor axis $a$.
  Contrarily to Fig.~\ref{fig_map_J_e_stacked}, we do not use here the quantity $\overline{a}$ because its definition depends on the particular resonance considered (Eq.~\ref{eq_modified_a}). However, because $e$ remains small, this choice does not change the interpretation of this figure taken from the run shown in Fig.~\ref{fig_passage_resonances_t_Lz}. We note that at time $T=$~21,600, the narrow second-order 3/5 SOR is able to significantly perturb the ring, in spite of the fact that the first-order 2/3 SOR is still expected to exert a non-negligible perturbation, amounting to 10\% of the eccentricity forced by the 3/5 SOR.  At time $T=$~112,000, we see the damping effect of collisions as the ring evolves between the 3/5 and 1/2 SORs. Note that 3$^{\rm rd}$-order resonances, for example the 4/7 SOR at $a$=1.45, have no effect on the migration.
  A movie generated from these snapshots is available online.
  }
  \label{fig_passage_resonances_a_e}
\end{figure*}

We have followed the migration of a colliding ring initially placed near the second-order 5/7 SOR. The Fig.~\ref{fig_passage_resonances_t_Lz} shows the time evolution of the particle angular momenta $L_z$. The crossing of each first or second-order SOR by the material results in a jump in $L_z$, superimposed to a general positive drift of $L_z$, i.e. a positive torque exerted by the mass anomaly on the disk.
The Fig~\ref{fig_passage_resonances_a_e} shows the evolution of the same system in the $(\overline{a},e)$ space. This figure confirms the expectation that each first- and second-order SOR excites the orbital eccentricities as predicted by Figs.~\ref{fig_num_vs_theo_scalings} and \ref{fig_map_J_e_stacked}, while collisions damp the eccentricities when the ring material evolves between resonances.
For the case $\mu=0.003$, Fig~\ref{fig_passage_resonances_a_e} shows that the characteristic timescale for the ring migration is some $10^{5}$ Chariklo's rotations, corresponding to some centuries, confirming the results obtained by the toy model of \cite{sicardy2019}. 

The elongations of bodies like Chariklo, Haumea or Quaoar cause stronger resonances than a mass anomaly (see  Figs.~7-9 of  Paper~I), and thus repel even more rapidly the ring material, with larger eccentricities. In that context, the weaker second-order 1/3 resonance may correspond to a protected zone where a ring may be confined.

\section{Effect of an outer satellite}
\label{sec_effect_satellite}

\begin{figure}
  \includegraphics[width=1\columnwidth]{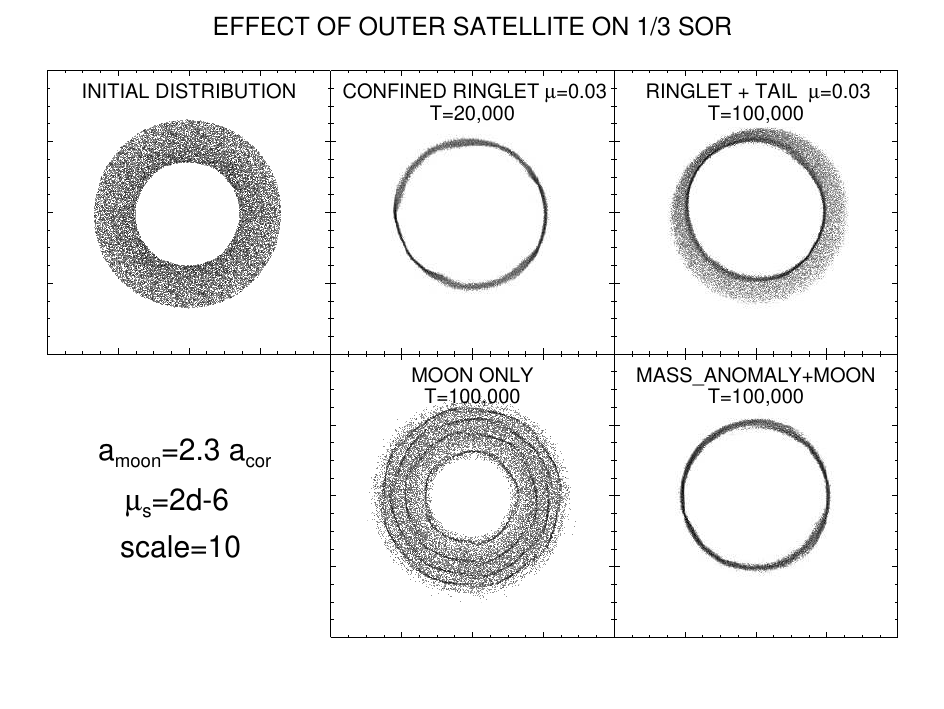}
  \caption{
  \textit{Upper row}: the long term evolution of a 1/3 SOR ringlet perturbed by a mass anomaly $\mu = 0.03$, using 30,000 particles with radius $R= 5 \times 10^{-4}$ (corresponds to 100~m for Chariklo rings). Although the ringlet maintains a well defined core, a slow outward leakage of particles appears, forming a faint tail. For better viewing, the width of the ring has been expanded by a factor of ten. 
 \textit{Lower left panel}: the effect of a satellite located at $2.3 \approx 1.1 a_{1/3}$, with a mass $\mu_{\rm s} = 2 \times 10^{-6}$ relative to the central body. Without the 1/3 SOR, the ring spreads inwards under the effects of spiral density perturbations forced by Lindblad resonances, shown here at $T$=100,000.
 \textit{Lower right panel}: when both the satellite and 1/3 SOR are present, the outward leakage of particles is prevented by the  8/7  Lindblad resonance with the satellite.
 Movies generated from these snapshots are available online.
 }
 \label{fig_effect_satellite}
\end{figure}

\begin{figure}
  \includegraphics[width=1\columnwidth]{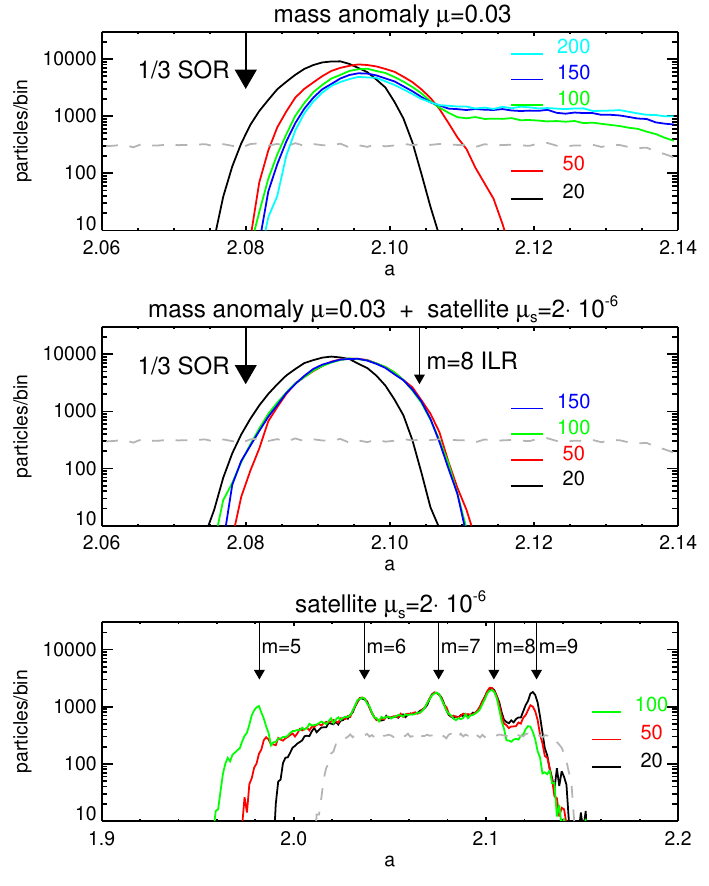}
  \caption{
  Comparison of the semimajor axis  distributions in the three simulations of Figs.~\ref{fig_effect_satellite}  and ~\ref{fig_effect_satellite_lz}. The upper frame is the run with mass anomaly, and the colors indicate profiles at various times $T=$~20,000 to 200,000. The dashed gray line is the initial distribution, and the arrow marks the 1/3 SOR location. Notice the growing tail of the distribution in profiles for $T \gtrsim$~100,000. In the middle frame, the simulation includes both the mass anomaly and a satellite. The run extends to $T=$~150,000 revolutions, with no leakage or spreading of the ringlet.  Bottom frame: simulation with a satellite alone, with arrows marking its inner Lindblad resonances. Notice the inward spreading of the ring and the gradual accumulation of material to the $m=5$ resonance, and the disappearance of the $m=9$ peak.
  }
\label{fig_effect_satellite_profiles}
\end{figure}

\begin{figure}
\centering
  \includegraphics[width=0.9\columnwidth]{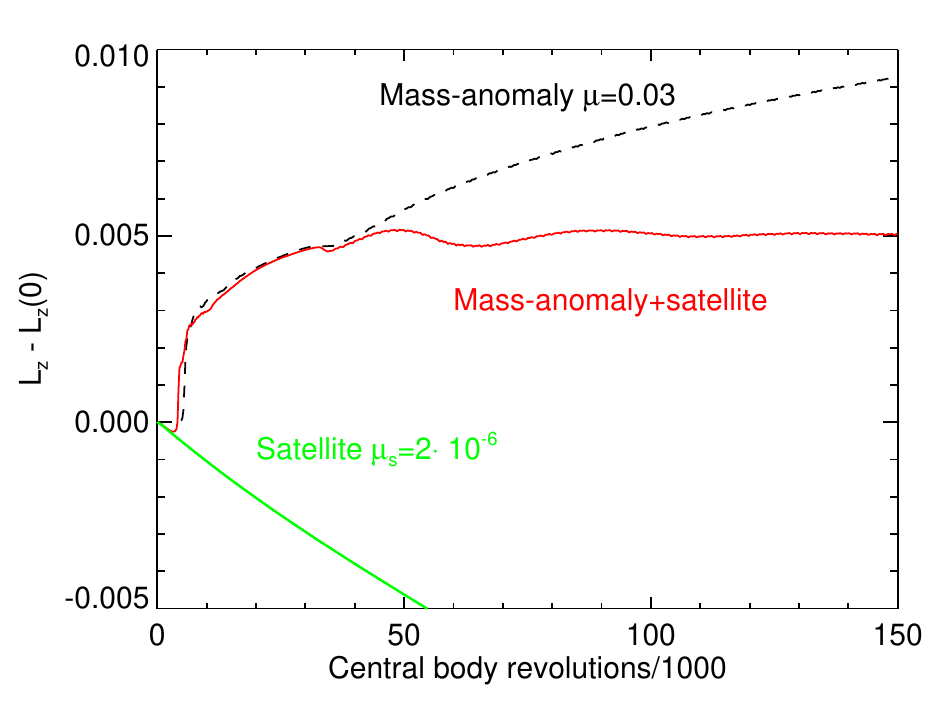}
  \caption{
  Evolution of mean angular momentum of the 1/3 SOR ring perturbed by a $\mu=0.03$ mass anomaly (black curve), by a $\mu_{\rm s}= 2 \times 10^{-6}$ satellite (green), or simultaneously by both (red). The snapshots of these simulations at $T=$~100,000 were displayed in Fig. \ref{fig_effect_satellite}.
  }
\label{fig_effect_satellite_lz}
\end{figure}

Even though the 1/3 SOR is efficient in confining a ring on short timescales, our simulations show a leakage of material outwards on the long term, see the upper rows of Figs.~\ref{fig_effect_satellite} and Figs.~\ref{fig_effect_satellite_profiles}, displaying the evolution in a simulation with $\mu=0.03$ up to $T=$~200,000. An eventual dispersal of the ringlet is unavoidable as the continuous torque exerted on the ringlet by the mass anomaly implies a gradual increase in its mean $L_z$, even after the rapid jump in $L_z$ associated with the resonance passage is over, see Fig.~\ref{fig_effect_satellite_lz}. 

The drift of the ringlet out of the resonance is not as rapid as the value of $dL_z/dt$ would indicate, since most of the angular momentum is carried out by particles pushed out from the resonance, while the core of the ringlet stays confined. However, even the ringlet core would eventually erode out. Estimated from the profiles in Fig.~\ref{fig_effect_satellite_profiles}, this might take about $10^6 - 10^7$ central body revolutions in the $\mu=0.03$ simulation, corresponding to about 800-8000 years in the Chariklo case. Without an active confinement, the particles pushed outside the resonance experience continuous viscous spreading. 

In order to avoid this leakage, a small satellite may be placed outside the ring, as illustrated in the lower row of Fig.~\ref{fig_effect_satellite}. This hypothetical satellite exerts a negative torque on the ring material through Inner Lindblad Resonances (ILR) $m/(m-1)$, where $m>0$. If acting alone, such a satellite will excite spiral density enhancements in its inner Lindblad resonances, and the associated torque will push the ring inward (lower left frame in Fig.~\ref{fig_effect_satellite} and the bottom row in Fig.~\ref{fig_effect_satellite_profiles}). If combined with a mass anomaly, the satellite may stop the outward leakage of particles from the ringlet, and lead to a steady state where the outward torque by 1/3 SOR is balanced by the satellite induced negative torque. The torque associated with a $m/(m-1)$ resonance is classically given by (see e.g. \cite{sicardy2019}
\begin{equation}
\Gamma_m = {\rm sign}(n_{\rm s}-n) 3 \pi^2 |m(m-1)^3| GM \Sigma_0 \epsilon'^2,
\label{eq_torque_LR}
\end{equation}
where $n_{\rm s}$ is the satellite mean motion, $\Sigma_0$ is the ring surface density and $\epsilon'$ is a dimensionless coefficient defined in  Eq.~17 of  Paper~I that measures the strength of the $m/(m-1)$ ILR. The coefficient $\epsilon'$ is proportional to $\mu_{\rm s}$, the mass of the satellite relative to the mass of the primary (not to be confused with the mass anomaly $\mu$ of the central body).

The satellite can halt the outward leakage of the ring if $|\Gamma_m|$ is larger than the viscous torque $\Gamma_\nu = 3\pi n_0 a_0^2 \nu \Sigma_0$. Using Eq.~\ref{eq:kvisc}, the condition $|\Gamma_m| \gtrsim \Gamma_\nu$ provides an order-of-magnitude estimation of the strength $\epsilon'$ (and thus on the satellite mass $\mu_{\rm s}$) necessary to balance the ring outward diffusion,
\begin{equation}
\epsilon' \gtrsim 
\sqrt{ \frac{ k_{\rm visc} \tau}{\pi m(m-1)^3} }
\left( \frac{R}{a_0} \right),
\label{eq_epsilon'_anti_leakage}
\end{equation}
where $a_0 \approx 2.08$ for the 1/3 SOR. Using the methodology of Paper~I, it can be shown that $\epsilon' \approx 0.6/m$ for $m$'s larger than a few times unity. From our simulations with a rebound coefficient $\epsilon_{\rm n} = 0.1$, we obtain $k_{\rm visc} \approx 3.5$, knowing that larger values of $k_{\rm visc}$ would increase that factor a bit without changing its order of magnitude. Using these values, the equation~\ref{eq_epsilon'_anti_leakage} can be re-expressed as 
\begin{equation}
\mu_{\rm s} \gtrsim 
0.8 \sqrt{\frac{|m| \tau}{|m-1|^3}} R,
\label{eq_satellite_mass_anti_leakage}
\end{equation}
when restricted to a 1/3 SOR ringlet. 

We apply this equation to the run shown in Fig.~\ref{fig_effect_satellite} where $m=8$, and $R= 5 \times 10^{-4}$.  
Inserting the peak optical depth at the ringlet core, $\tau =0.15$, would require $\mu_{\rm s} \gtrsim 3 \times 10^{-5}$. However, since the ringlet itself is confined by the 1/3 SOR, it is more relevant to use the $\tau$ of the tail in this estimate, $\tau \approx 0.01$, which indicates $\mu_{\rm s} \gtrsim 7 \times 10^{-6}$. This is of similar order of magnitude with the result of Fig.~\ref{fig_effect_satellite}, where a  satellite with mass $\mu_{\rm s} = 2 \times 10^{-6}$ prevents the viscous spreading of the ring.  Note that the $m$-number has no special role in preventing the leaking:
depending on mass and distance of the exterior satellite, a balance could be achieved at an ILR with a different $m$.

  \section{Self-gravity}
  \label{sec_sg}
The simulations presented so far have not included mutual gravity between ring particles, but have concentrated on the collisional confinement at the resonances.  However, self-gravity is expected to have significant influence on the ring dynamics. In low density rings the gravitational scattering in binary encounters enhances the steady-state velocity dispersion and thereby the ring viscosity, while for larger densities the continuously forming and dissolving self-gravity wakes dominate the dynamics \citep{salo1992, salo1995}.  The
wakes increase the viscosity significantly, both via gravitational torques and due to increased velocity dispersion associated with wake motions \citep{daisaka2000}.  Finally, for sufficiently large planetocentric distances, the particles start to collect to gravity-bound aggregates. Although a fully realistic (see below) treatment of self-gravity would require much larger $N$ than achievable in our current simulations, we here briefly address how self-gravity might affect the ring evolution, and in particular whether the confinement
at 1/3 SOR could still work.

\subsection{Scaling with the $r_{\rm h}$ parameter}
\label{sec_rh}
The importance of self-gravity is conveniently described by the dimensionless parameter $r_{\rm h}$  \citep{ohtsuki1993, salo2018}, that compares the size of particle pair's gravitational Hill radius to their physical sizes. For a pair of identical particles at distance $a$ from
a spherical  central body with radius $R_{\rm B}$,
\begin{equation}
r_{\rm h} = \frac{R_{\rm H}}{2R} = \left(\frac{\rho}{12\rho_{\rm B}}\right)^{\frac{1}{3}} \left(\frac{a}{R_{\rm B}}\right),
\label{eq:rhdef}
\end{equation}
\noindent 
where $\rho$ and $\rho_{\rm B}$ are the bulk densities of the particles and  central body, respectively.  Here $R_{\rm H}=(2M_p/3M_{\rm B})^{1/3}a$ is the radius of the Hill-sphere of two particles with masses $M_p$, inside which the pair's mutual gravity dominates over the tidal pull from the central body at distance $a$. When $r_{\rm h}$ decreases, the particle pair extends more and more out from its Hill-sphere: $r_{\rm h} = 0$ corresponds to the non-gravitating case, while for  $r_{\rm h}=1$, the net attraction between two synchronously rotating, radially aligned ring particles in contact equals the disruptive tidal force. The classical Roche limit, $a/R_{\rm B} \le 2.456 ~(\rho_{\rm B}/\rho)^{1/3}$ for the tidal disruption of a fluid body corresponds to $r_{\rm h} \le 1.072$.

Both the influence of binary encounters and self-gravity wakes can be written in terms of $r_{\rm h}$, in addition to $\tau$ and $nR$ which alone were sufficient to describe the state of a non-gravitating ring for a given coefficient of restitution.  The velocity dispersion maintained by encounters is comparable to the two-body escape velocity, $v_{\rm esc}= \sqrt{2GM_p/R}$. In terms of $r_{\rm h}$, this velocity dispersion is
\begin{equation}
  \frac{c_{\rm esc}}{n R}= 4.9 {r_{\rm h}}^{3/2}.
\end{equation}
\noindent
For $r_{\rm h} \gtrsim 0.7$, $c_{\rm esc}$ exceeds the typical dispersion maintained by impacts alone, $c_{\rm imp} \approx (2-3) nR$.  At small $\tau$ when self-gravity wakes are weak, the viscosity is enhanced due to encounters roughly by a factor $(c_{\rm esc}/c_{\rm imp})^2$ compared to the value given in Eq.~\ref{eq:kvisc}. For $r_{\rm h} \sim 1$,  this corresponds to a factor of $\sim 4$ increase in viscosity. A same enhancement would be obtained by  using a particle size increased by a factor of $\sim 2$ in a non-gravitating simulation. 

Similarly, the Toomre critical wavelength and velocity dispersion can be written in terms of $r_{\rm h}$ as \citep[see,][]{salo2018}
\begin{equation}
\frac{\lambda_{\rm cr}}{R} = 48 \pi \ \tau {r_{\rm h}}^3,
\label{eq:lcrr}
\end{equation}
\begin{equation}
\frac{c_{\rm cr}}{n R}  = 12.8 \ \tau {r_{\rm h}}^3.
 \label{eq:cr}
\end{equation}
\noindent The wake structure, with a typical radial spacing $\sim \lambda_{\rm cr}$ starts to emerge whenever the radial velocity dispersion maintained by impacts drops below $(2-3) c_{\rm cr}$, corresponding to $\tau {r_{\rm h}}^3 \gtrsim 0.1$. Strong wakes imply substantially increased viscosity, both due to gravitational torques exerted by the wakes and due to increased random motions. The standard formula for the gravitational viscosity reads \citep{daisaka2000}
\begin{equation}
\nu_{\rm grav} \approx 190 {r_{\rm h}}^{11} \tau^2 nR^2,
\end{equation}
and including the wake motions, leads to total $\nu_{\rm tot} \approx 2\nu_{\rm grav}$. Comparing to Eq.~\ref{eq:kvisc}, this implies an enhanced viscosity by a factor $\sim 100 \ \tau {r_{\rm h}}^{11}$ over a non-gravitational system (see also Fig. B1 in \cite{salo2025}). 

For $r_{\rm h}$ approaching unity, the wake structure becomes increasingly clumpy, and for $r_{\rm h} \gtrsim 1.1-1.2$ the wakes degrade to semi-permanent particle aggregates \citep{salo1995,karjalainen2004}. Accretion takes place also in low density rings via pairwise accumulation of particles. See Fig. 16.24 in \cite{salo2018} for an illustration of various parameter domains dominated by impacts, encounters, self-gravity wakes, and gravitational accretion.

\subsection{Self-gravitating simulations}

\subsubsection{Accretion boundary}
Inserting the values of Table~\ref{tab_param_cha} into Eq. \ref{eq:rhdef} implies that for the Chariklo ring system
\begin{equation}
r_{\rm h} = 0.697 \left(\frac{\rho}{900 {\rm~kg~m}^{-3}}\right)^{1/3} 
\frac{a}{a_{\rm cr}} =1.45\ \left(\frac{\rho}{900 {\rm~kg~m}^{-3}}\right)^{1/3} 
\frac{a}{a_{1/3}},
\label{eq:rh}
\end{equation}
so that the 1/3 SOR distance corresponds to $r_{\rm h} \gtrsim 1$ whenever $\rho \gtrsim 300$~kg~m$^{-3}$.  Thus, even for
relatively under-dense particles with $\rho=200-300$~kg~m$^{-3}$, the C1R ring region should be strongly affected by self-gravity.

\begin{figure*}
\includegraphics[width=2\columnwidth]{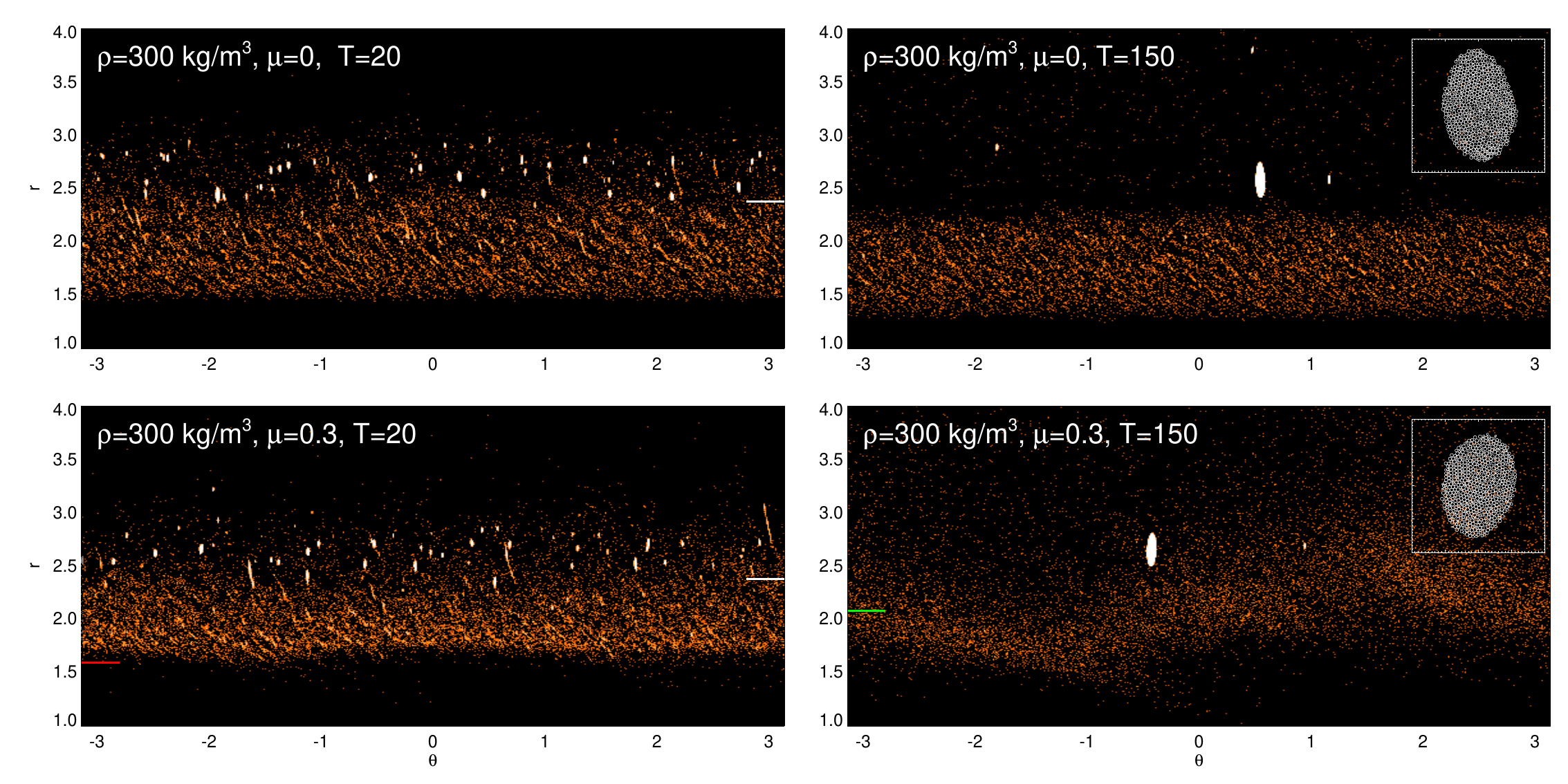} 
\caption{
  Self-gravitating simulations with $R=0.005$ (corresponding to a physical radius of 1~km in the Chariklo case), assuming particles with bulk density $\rho=300$~kg~m$^{-3}$. The initial distribution extends $r= 1.55 - 2.8$, which corresponds to the range  $r_{\rm h}=0.75-1.35$. With $N=24,000$ particles the initial $\tau_0=0.11$. The upper row shows two snapshots of the simulation with no mass
  anomaly, while the lower row shows a simulation with a $\mu=0.3$ mass anomaly. The location of the 1/3 SOR ($a=2.08$) corresponds to $r_{\rm h} \approx 1.0$. In the $T=20$ frames, the tick marks on the right vertical axis indicate the radii where $r_{\rm h}=1.15$. The red and green solid lines in the $\mu=0.3$ frames indicate the locations of 1/2 (left) and 1/3 (right) SOR.    The inserts in the $T=150$ frames zoom-in to the largest aggregate formed during the run, covering a $0.4 \times 0.4$ region with a correct aspect ration. The slice of the aggregate through the $z=0$ plane is shown.
  }
\label{fig_sg1}
\end{figure*}

Figure \ref{fig_sg1} compares two simulations starting with a wide initial ring around the 1/3 SOR, spanning the range $a= 1.55 - 2.80$; for the adopted bulk density of particles ($\rho=300$~kg~m$^{-3}$) this corresponds to $r_{\rm h} = 0.75 - 1.35$. In the first simulation (upper row) a spherical central body is used, while the second (lower row) uses $\mu=0.3$. A large particle radius $R=0.005$ (corresponding to 1~km in Chariklo's case) is adopted, yielding an initial optical depth $\tau_0=0.11$, which should be sufficiently large for self-gravity wakes to become discernible. A large $\tau$ also speeds up the formation of gravitational aggregates at large distances. Indeed, in both simulations particle aggregates start to form rapidly, within first
few ring orbital periods, in the region $r_{\rm h} \gtrsim 1.15$ (marked by the solid line on the right axis of the $T=20$ frames).  Inside this distance self-gravity wakes form, inclined by $\sim 20^\circ$ with respect to the tangential direction and with radial spacing $\sim 0.1$, consistent with Eq. \ref{eq:lcrr}. There are also a few elongated streaks visible at the border zone between the wake and accretion regions: these are formed by particle aggregates recently destroyed by tidal forces or due impacts. Eventually, the aggregates manage to merge, and in the end of both runs at $T=150$, about $95\%$ of the particles beyond $r_{\rm h}=1.15$ have collected into a single aggregate. A significant viscous spreading is revealed by the inward motion of the ring inner edge in the $\mu=0$ simulation.

For the simulation with a mass anomaly, a large $\mu$ was chosen, in order to make the resonance perturbations significant in spite of the large $R$ and the very short duration of the run. Indeed, the particles are gradually pushed outward due to resonance torques: in
the frame at $T=20$, a weak $m=2$ undulation is visible, related to the $1/2$ SOR at $a=1.59$. Conversely, the large amplitude $m=1$ pattern at the $T=150$ snapshot is related to $1/3$ SOR. However, the ring viscosity is much too large to allow an efficient resonance confinement to take
place. Also note that self-gravity wakes and particle clumps are totally absent in the perturbed region at the $T=150$ frame (lower right panel of Fig.~\ref{fig_sg1}).

\begin{figure*}
\includegraphics[width=2\columnwidth]{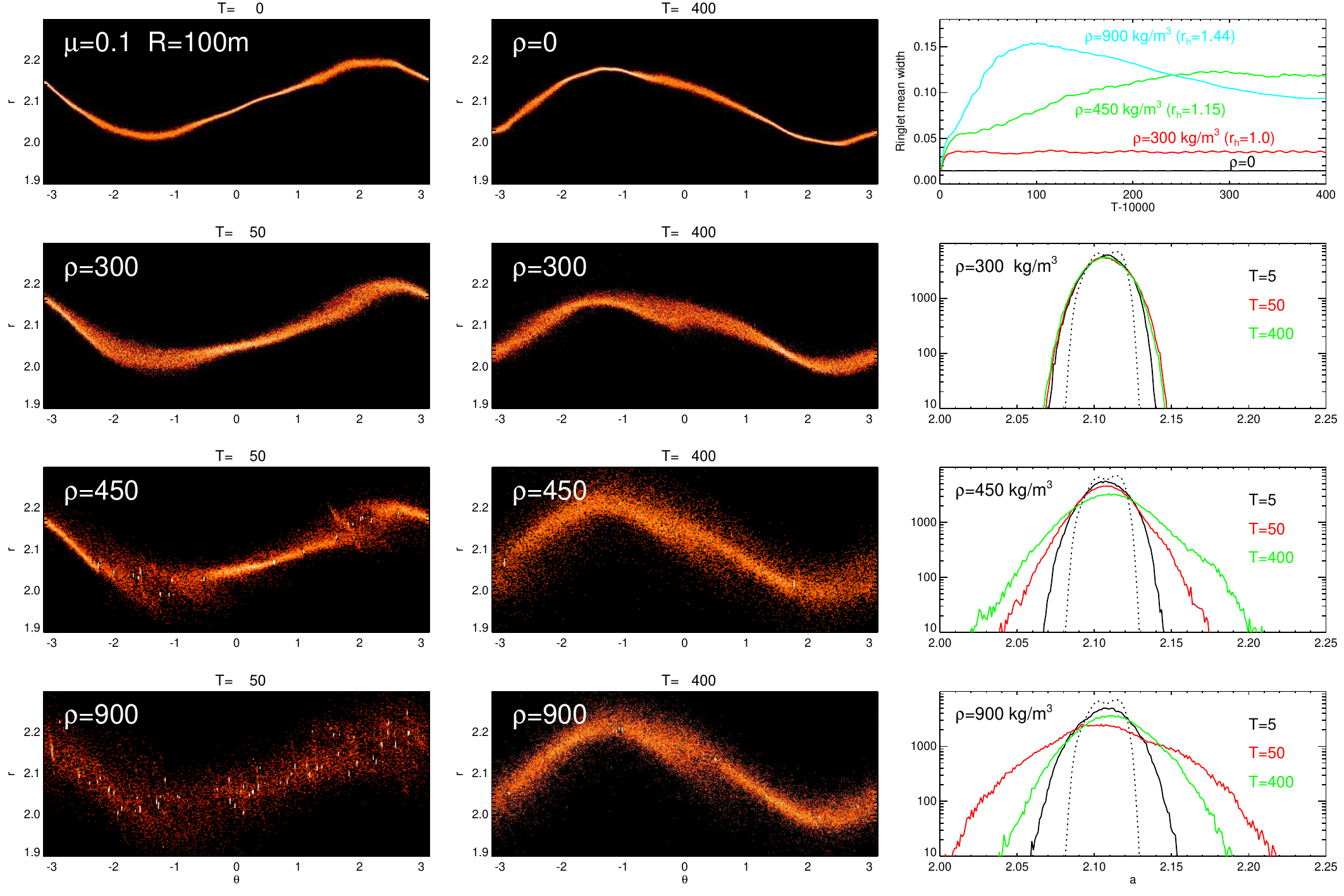}
\caption{
 Self-gravitating simulations of $N$=30,000 particles with different bulk densities. Simulations continue from the ringlet stage (at $T=10,000$) of the non-gravitating simulation of Fig.~\ref{fig_threshold} with $\mu=0.1$, $R=0.0005$ (100~m in Chariklo's case): the snapshot of the initial state is shown in the upper left frame. The other frames display self-gravitating simulations with $\rho$= 300, 450, and 900~kg~m$^{-3}$, corresponding to $r_{\rm h}=1.0, 1.15$, and $1.44$, respectively at the distance of the 1/3 SOR. Times are counted from the beginning of the self-gravitating simulation.
The upper right frame displays the evolution of the ringlet mean width, $W=\sqrt{12 <(r-r_{\rm fit})^2>}$, where $r_{\rm fit}$ denotes the fitted ringlet central line at the azimuth of the particle and the averaging is over all particles (for a uniform particle distribution this formula with the factor 12 recovers the full width of the ringlet). The remaining frames in right show the semi-major axis distribution at different instants for the three self-gravitating runs.
}
\label{fig_sg2}
\end{figure*}

\subsubsection{Resonance confinement}
\label{sec_sg_confinement}

The condition derived in Section \ref{sec_scaling} for the resonance confinement implies $\mu_{\rm thresh}  \propto \sqrt{\nu}$. To check whether the resonance confinement is possible in the presence of self-gravity in spite of the viscosity enhancement, we added particle gravity to a non-gravitating simulation that has already formed a confined ringlet. We selected as a starting point the run shown in the upper-left corner of Fig. ~\ref{fig_threshold}, corresponding to  $\mu=0.1$ and $R=0.0005$ (100~m in Chariklo's case), For this value of $\mu$, even a 4-fold increase in $R$ (i.e. a factor of 16 in viscosity) still permits a confinement in the non-gravitating case. We note that this radius $R$ is 10-times smaller than in the examples of Fig. ~\ref{fig_sg1}.  Three simulations with $\rho=300, 450$, and 900~kg~m$^{-3}$, corresponding to $r_{\rm h}=1.0, 1.15$, and 1.44, respectively, started from the ringlet stage at $T=10,000$. The time evolution of particle distributions with various $\rho$'s is followed in Fig. ~\ref{fig_sg2}. The first and second columns display polar snapshots at $T=50$ (about 17 ring revolution periods after the inclusion of self-gravity), and at $T=400$ (the end of the self-gravitating run), while the right column shows the time evolution of the ringlet mean width (uppermost frame) and the semi-major axis distributions.

With $\rho=300$~kg~m$^{-3}$ the ringlet remains well-confined with sharp inner and outer boundaries. Nevertheless, its width has become about two-fold compared to the non-gravitating case.  This increase in the width takes place rapidly during the first few ring revolutions, accompanied by a similar increase in the velocity dispersion. During later evolution, the width of the ring as well as the dispersion of semi-major axis is practically constant.  With larger bulk  densities a rapid formation of particle aggregates takes place.  At $T=50$ with $\rho=450$~kg~m$^{-3}$, the ringlet has developed several azimuthal gaps, connected to the largest individual aggregates which scatter particles out from the ringlet core.  For $\rho=900$~kg~m$^{-3}$, the whole ringlet has become quite fuzzy.  Nevertheless, after the initial evolution the ringlet spreading seems to stop ($\rho=450$~kg~m$^{-3}$) or even become reversed ($\rho=900$~kg~m$^{-3}$).

\subsubsection{Cyclic formation and destruction of aggregates}
 \label{sec_sg_cyclic}

\begin{figure*}
\includegraphics[width=2.1\columnwidth]{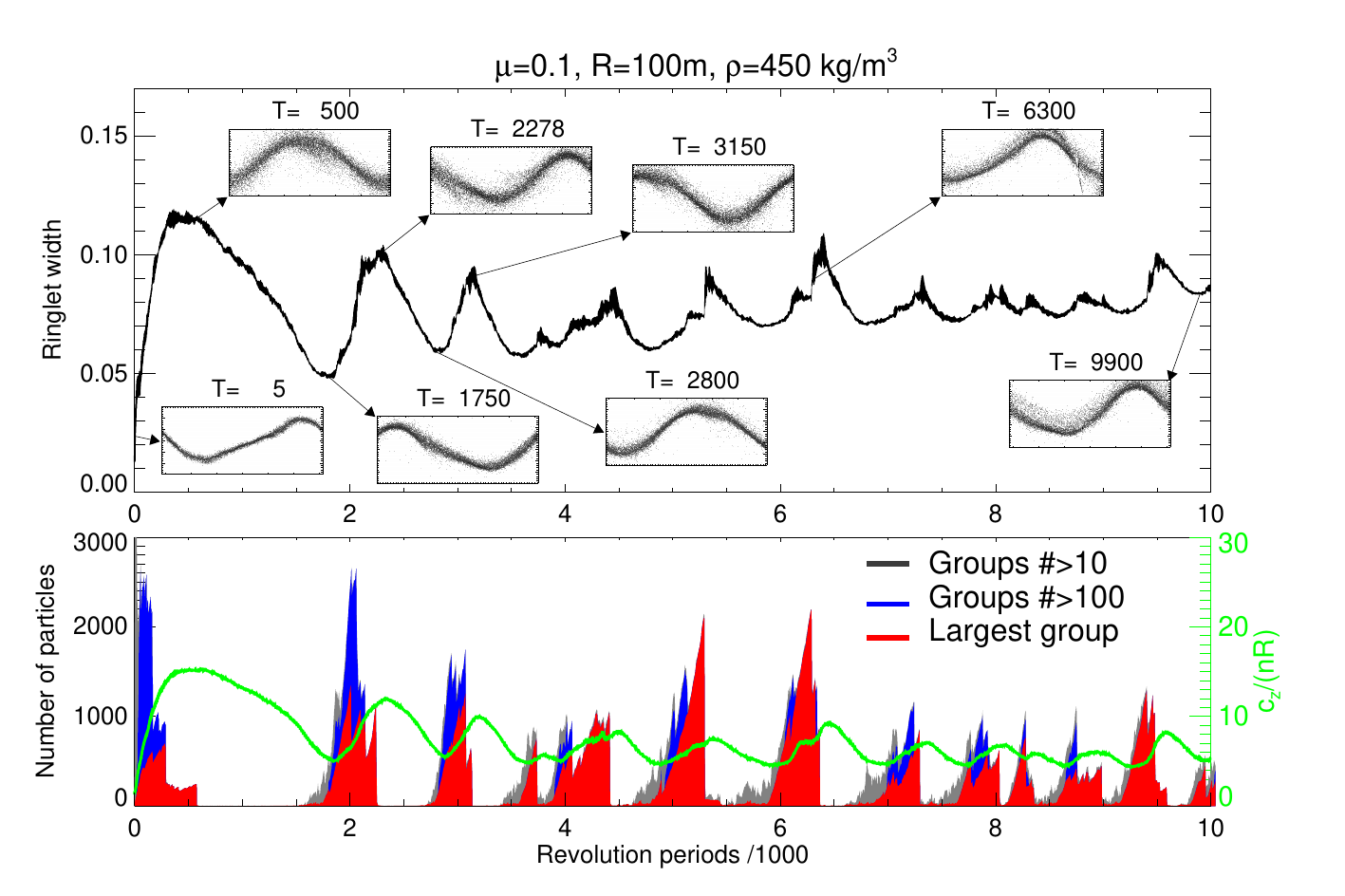} 
\caption{
Self-gravitating simulations with $\mu=0.1$, $R=0.0005$ (100~m in Chariklo's case), assuming particles with bulk density $\rho=450$~kg~m$^{-3}$. The initial particle distribution is similar to that in Fig. \ref{fig_sg2}, except that $\tau$ has been reduced by a factor of two by selecting only every other particle from the confined ringlet in the non-gravitating simulation at $T=10,000$. Upper frame displays the evolution of the ringlet mean width, together with snapshot corresponding to various width maxima and minima; the snapshot at $T=6300$ displays the system just after tidal breakup of the largest aggregate.  The lower frame displays the number of particles in the largest aggregate (red) and the total number of particles in groups with more than hundred (blue) and  ten particles (gray; the total number of particles $N=15,000$). Also shown as a green curve is the vertical dispersion $c_z$ (labels in the right).
}
\label{fig_sg3}
\end{figure*}

Based on Fig. ~\ref{fig_sg2} it seems plausible that a ringlet can remain confined at the 1/3 SOR for values $\rho \le 900$~kg~m$^{-3}$. However, due to the short duration of the simulation, it is unclear whether an actual steady-state has been achieved in the runs with $\rho=450$~kg~m$^{-3}$ or $900$~kg~m$^{-3}$.  Therefore, another simulation with $\rho=450$~kg~m$^{-3}$ was conducted, now extended over 20-times longer duration but reducing the particle number by one half ($N=15,000$) to speed-up the calculation.  Fig. ~\ref{fig_sg3} shows the evolution of ringlet width
and velocity dispersion\footnote{ We use $c_z$ as an easy-to-calculate proxy of the local velocity dispersion since it is not directly affected by the systematic velocities induced by resonance perturbations. At the same time $c_z$ is effectively coupled to the planar random velocities via impacts.}, and also quantifies the amount of particles in aggregates, separately in the small ($10<N_{group}<100$; gray color) and large groups ($N_{group}>100$; blue), and in the largest aggregate (red).  

The initial evolution in the simulation is practically similar to the corresponding run in Fig. ~\ref{fig_sg2}: the velocity dispersion  ($c_z \sim Rn$) is initially less than the escape velocity of individual particles ($c_{esc} \sim 6 Rn$) and therefore small particle groups are rapidly formed. However, the gravitational stirring by the growing groups heats the system, soon preventing further pairwise accretion. The already formed groups gradually grow through merging (and occasionally destroy each other in impacts), until at $T\sim 200$, the system is dominated by one large aggregate, with mass of the order of 4\% of the total ringlet mass ( i.e. involving $\sim 600$ particles), exceeding the total mass in other aggregates. At $T\sim 300$ this large aggregate looses half of its mass by tidal leakage and at $T=500$ it breaks completely. When this happens, the system starts to cool down due to collisional dissipation as there is no more gravitational stirring by the aggregates. 

Interestingly, the aggregate formation/destruction appears to be a self-regulating process: once the system has cooled down sufficiently ($c_z \sim 5Rn$, at $T \approx 1750$), the accretion starts again, going through the formation phase of increasingly large particle groups, until at $T \approx 2300$ the last large aggregate is again tidally destroyed, and the cooling phase starts again. This cycle repeats throughout the simulation, though with somewhat diminished variations. The alternating cycle is also visible in the snapshots of the system, the ringlet appearance varying from sharper to fuzzier, depending of whether aggregates are present or not (see the inserts in Fig. ~\ref{fig_sg3}: the individual aggregates are too small to be discernible in the plots, except indirectly in the snapshot for $T=6300$, displaying the system just after a very large particle aggregate with 2000 particles has been tidally shredded, the debris being still visible). 

The simulation of Fig.~\ref{fig_sg3} suggests that quite an interesting type of behavior is possible near the accretion boundary at $r_{\rm h} \sim 1.1-1.2$. To check that the observed tidal disruption of large aggregates is not a numerical artifact, but indeed due to the perturbation, an additional simulation was conducted. It started from the above simulation at $T=6250$, shortly before the break-up of the big aggregate, and used a time step reduced to one-half (1/1200 of the ring orbital period). Also in this case the aggregate broke up, although a bit later than in the original run. In both runs the break-up happened via gradual stretching in the radial direction. On the other hand, another experiment starting from $T=6250$, where the single large aggregate was isolated and all other particles removed, led to a stable aggregate (surviving to the end of the run lasting to $T=7000$). This suggest that the tidal break-up is induced by impacts from particles perturbed by the resonance, rather than some secular instability due to errors accumulated during the orbital integration of the aggregate.

\section{Discussion}

Our main result is that a mass anomaly $\mu$ leads to the confinement of ring material near the 1/3 SOR, provided that the strength of the resonance overcomes the spreading effect of collisions, see Eq.~\ref{eq_threshold} and  Fig.~\ref{fig_threshold}. In analogy with Saturn's dense rings, we may assume that the dynamics of the dense rings observed so far is dominated by large particles with radius of the order of a meter \citep{cuzzi2018}. Using for instance a corotation radius $a_{\rm cor} \sim 200$~km in the Chariklo case (Table~\ref{tab_param_cha}), we obtain $R \sim 5 \times 10^{-6}$ in Eq.~\ref{eq_threshold_bis}. Considering that $\tau_{\rm Q1R} \sim 1$, this gives a threshold value $\mu \gtrsim 10^{-3}$ that permits ring confinement. If the mass anomaly $\mu$ is caused by a mountain of height $h$ at the surface of a body with radius $R_{\rm ref}$, then $\mu \sim 0.5(h/R_{\rm ref})^{3}$. Using $R_{\rm ref}=115$~km for Chariklo, then $\mu \sim 10^{-3}$ corresponds to $h \sim$~10~km, a plausible value when compared with other small objects of the Solar System.

Eq.~\ref{eq_threshold_bis} shows that the threshold value $\mu$ is proportional to $R$, which is the particle size normalized to $a_{\rm cor}$. Consequently, for a given particle size in meters, $\mu$ scales like $a_{\rm cor}^{-1}$. Since Haumea's and Quaoar's corotation radii are respectively six and ten times greater than for Chariklo ( Table~2 of  Paper~I), the threshold value of $\mu$ is of the order of $10^{-4}$ for these two bodies.

It remains to be seen if these objects can sustains such mass anomalies. They can be compared with various small bodies of the Solar System that exhibit large topographic features\footnote{https://en.wikipedia.org/wiki/List\_of\_tallest\_mountains\_in\_the\_\\Solar\_System}. For instance Ceres ($R_{\rm ref} \sim 470$~km) has mountains with $h/R_{\rm ref} \sim 8.5 \times 10^{-3}$, while Vesta ($R_{\rm ref} \sim 270$~km) has features with $h/R_{\rm ref} \sim 8.4 \times 10^{-2}$. This corresponds to $\mu= 3 \times 10^{-7}$ and $\mu= 3 \times 10^{-4}$, respectively. The large Trans-Neptunian Object (307261) 2002 MS$_4$ ($R_{\rm ref} \sim 400$~km) possesses a large depression with depth $\sim 45$~km and an extension of some 320~km, corresponding to $\mu$ of a few times $-10^{-2}$ \citep{rommel2023}. 

As noted in subsection \ref{sec_triaxial_vs_mass_anom}, while Chariklo's, Haumea's and Quaoar's main rings orbit close to the 1/3 SOR, Quaoar's fainter ring Q2R is close to the second-order 5/7 SOR. This is the starting point of the simulation shown in Fig.~\ref{fig_passage_resonances_a_e}. However, this run uses Chariklo's parameters, and the ring is rapidly evacuated from this resonance due to the large effects of the nearby 3/4 and 2/3 SORs. In the case of Quaoar, the 5/7 SOR is better isolated from the 3/4 and 2/3 SORs ( Fig.~9 of  Paper~I), and future numerical simulations could test the ability of the 5/7 SOR to confine a ring. This said, we note in Fig.~\ref{fig_passage_resonances_a_e} that the material is temporarily trapped in another second-order SOR corresponding to $n/\Omega_{\rm B}= 3/5$, showing that the 1/3 resonance is not the only one that can confine a ring.

Our simulations show that the ring confinement is accompanied by the excitations of various free Lindblad modes with azimuthal wave numbers $m$ (Fig.~\ref{fig_mode_analysis}). This was observed most strikingly in our ``best" simulation with $\mu=0.003$ and $R=25$~m, where the resonance excitation was eventually divided between normal modes with several values of $m$ (Fig.~\ref{fig_mode_analysis}). In simulations with larger $\mu$'s and $R'$, signs of similar excitations were seen, but not all normal modes were present. The presence of such modes could be revealed using multi-chord stellar occultations by providing accurate  orbital radii and ring widths at various longitudes. Unfortunately, even for the best-observed  occultations by Chariklo's system, the azimuthal sampling of the rings is still too coarse, Chariklo's center  cannot be pinned down accurately, and the ring precession rate $\dot{\varpi}$ is poorly constrained. This prevents any accurate analysis of these modes. Hints for the presence of $m=-1$ and $m=-2$ Lindblad modes in Chariklo's main ring C1R has been reported by \cite{gomes2025}. Although encouraging, these detections still require confirmation through more detailed observations.

As discussed in Section~\ref{sec_effect_satellite} and shown in Fig.~\ref{fig_effect_satellite}, the 1/3 SOR confinement by a mass anomaly alone is not sufficient to explain the long term presence of a ring. A small satellite is required to balance the outward viscous spreading of the confined ringlet. Eq.~\ref{eq_satellite_mass_anti_leakage} shows that the mass of this satellite is proportional to the size of the ring particles $R$. 
Assuming meter-sized particles, this yields $R = 5 \times 10^{-6}$ in the Chariklo case, which is 100~times smaller that the radius used for the run shown in Fig.~\ref{fig_effect_satellite}. Eq.~\ref{eq_satellite_mass_anti_leakage} then indicates that a satellite with mass $\mu_{\rm s} \sim 10^{-7}$ can stabilize the outwards spreading of a 1/3 SOR ringlet. This value is comparable to the masses obtained by \cite{sickafoose2024} -- several times $10^{-6}$ -- estimated from their numerical simulations of Chariklo's rings, taking particle radii of a few meters confined by various inner and outer Lindblad resonances.

The value $\mu_{\rm s} = 10^{-7}$ corresponds to an icy moonlet as small as 0.5~km in radius in the case of Chariklo. The presence of such small satellites can be expected as Chariklo's rings are close to the classical Roche limit, allowing the a mixture of satellites and rings in the 1/3 SOR region. The same is true for Haumea's and Quaoar's rings \citep{hedman2023}. We note from Eq.~\ref{eq_satellite_mass_anti_leakage} that again $\mu_{\rm s} \propto a_{\rm cor}^{-1}$. Thus, for Haumea and Quaoar, the threshold values of $\mu_{\rm s}$ are $\sim 1.5 \times 10^{-8}$ and $\sim 10^{-8}$, corresponding to icy moonlets of radii of the order of 2~km and 1~km, respectively.

The numerical estimates of the various threshold values for $\mu$ and $\mu_{\rm s}$ are based on non-gravitating simulations. However, for the range of plausible bulk densities of ring particles,  $\rho= 300-900$~kg~m$^{-3}$, self-gravity can be expected to have a significant effect in Chariklo's rings.  In terms of the Hill dimensionless parameter $r_{\rm h}$  (see Section~\ref{sec_sg}), this range of densities corresponds to $r_{\rm h}=1.0-1.44$. This implies that when applying Eq. \ref{eq_threshold_bis}, the viscosity enhancement due to gravitational encounters ($\tau \lesssim 0.1$) and/or self-gravity wakes ($\tau \gtrsim 0.1$) are both important. The enhancement due to gravitational encounters was estimated to be of the order of 5-10, while wakes can lead to even 100-fold viscosity when $r_{\rm h} \sim 1$. In terms of threshold value for $\mu$ which is proportional to square-root of viscosity,  such enhanced viscosities  would require roughly 3 and 10 times larger minimum size for the mass anomaly in order to obtain confinement, or similarly, a smaller maximum particle size for a given value of $\mu$.

The self-gravitating simulations of Section~\ref{sec_sg} indicate that resonantly confined ringlets are surprisingly resilient against gravitational accretion, whereas non-perturbed rings experience rapid accretion once $r_{\rm h} \gtrsim 1.15$.   In our self-gravitating experiments the starting point was taken from a confined ringlet formed in a non-gravitating simulation: after switching-on self-gravity the ringlet settled into a more diffuse, dynamically hotter state, while still remaining confined. During the initial evolution, when the velocity dispersion was still small, aggregates rapidly formed, but were eventually destroyed via tidal stretching.  Near the accretion limit $r_{\rm h}=1.15$ a quite interesting cyclic behavior was seen, with the system alternating between states where aggregates were abundant or nearly absent.

The low $\tau$ simulations of Section \ref{sec_sg} covered only the influence of gravitational encounters and particle accretion, as the current particle number is far too small for realistic modeling of self-gravity wakes: Fig. \ref{fig_sg1} was merely an illustration of the regimes where wakes/accretion take place, using a very large $R$. Indeed, a realistic modeling of a dense narrow ring with self-gravity wake structure would require $R << \lambda_{\rm cr} << W$, whereas in the non-gravitating simulations it is sufficient to have $R << W$, where $\lambda_{\rm cr}$ is the Toomre critical wavelength and $W$ is the ring width.  This extra intermediate length scale implies at least a factor of ten smaller particles, increasing $N$ by factor of at least 100 for a fixed $\tau$ (in fact, near Roche limit, $\lambda_{\rm cr}/R \sim 100$, indicating still larger separation of scales). Moreover, the optical depth should be of the order of 0.1-1, further increasing $N$. Thus at least $10^6-10^7$ particles would be needed, instead of the few tens of thousands employed in the current simulations\footnote{This is about the number of particles, $N=10^6 - 5\times 10^6$, recently employed by \cite{torii2024} in their global simulations of embedded satellites in Saturn's rings.}.

\section{Conclusions}

In connection with the analytical calculations of Paper~I, we have investigated the effects of spin-orbit Resonances (SORs) on a collisional ring surrounding an irregular body, using fully 3D collisional simulations. Our main conclusion is that the 1/3 SOR between the ring and the central body is efficient in confining ring material, while a small outer satellite may prevent the long-term outward spreading of material. More specifically, our main results are:
\begin{enumerate}
\item
Our simulations confirm the theoretical expectation that only first and second-order SORs effectively disturb a collisional disk.
\item
We can reproduce the peak eccentricity and the resonance width expected from our analytical calculations (Figs.~\ref{fig_num_vs_theo_scalings}, \ref{fig_map_J_e_snapshots_phases_I_II_III} and \ref{fig_passage_resonances_a_e}). This validates the use of the $(\overline{a},e)$ space to study the ring evolution (Fig.~\ref{fig_map_J_e_snapshots_phases_I_II_III}).
\item
The dense mesh of first and second-order resonances around Chariklo, Haumea and Quaoar excite large eccentricities and rapidly clears up a colliding disk up to the 1/2 resonance (Figs.~\ref{fig_passage_resonances_t_Lz}-\ref{fig_passage_resonances_a_e}). In that context, the outermost second-order 1/3 SOR forced by a mass anomaly is a ``quieter place", immune from the interactions with the inner SORs (Fig.~\ref{fig_num_vs_theo_e}).
\item
The 1/3 SOR first forces a self-intersecting streamline (Figs.~\ref{fig_map_J_e_snapshots_phases_I_II_III} and \ref{fig_wide_initial_distribution}) that gathers the background material of the disk into a ringlet. The self-intersection problem is avoided by the fact that the eccentricity excitation caused by the 1/3 SOR is transferred into superposed, non-self-intersecting free Lindblad modes, which eventually leads to the ring confinement (Figs.~\ref{fig_map_J_e_snapshots_phases_I_II_III} and \ref{fig_mode_analysis}).
\item
We obtain a condition for the confinement of a ringlet at the 1/3 SOR. It expresses the fact that the collisional viscous spreading is counteracted by the 1/3 SOR confinement effect. This condition, $4 \times 10^{-5} {\mu}^2 \gtrsim \tau R^2$, relates the optical depth $\tau$, the radius $R$ of the particles normalized to the radius of the synchronous orbit, and the value $\mu$ of the mass anomaly, see Eq.~\ref{eq_threshold_bis} and Fig.~\ref{fig_threshold}. Application to Chariklo shows that a mass anomaly $\mu \gtrsim 10^{-3}$ can confine material at the 1/3 SOR, which corresponds to typical topographic features (mountains or craters) of the order of 10~km.
\item 
A long term outward viscous spreading of the 1/3 SOR ringlet is observed in our simulations. We find that this leakage can be prevented through Lindblad resonances raised by an outer moonlet of mass $\mu_{\rm s} \gtrsim 10^{-7}$ (Fig.~\ref{fig_effect_satellite}) relative to Chariklo, corresponding to a sub-km moonlet. However, it is important to note that even in this case, the ring confinement is still ensured by the 1/3 SOR.
\item 
These results can be applied to Haumea's and Quaoar's rings by noting that the threshold values of $\mu$ and $\mu_{\rm s}$ mentioned above scale as the inverse of the radius of the synchronous orbit $a_{\rm cor}$.
Therefore, the threshold values of $\mu$ necessary to confine a 1/3 SOR ring are $\sim 10^{-4}$ for these two bodies. The threshold value of the satellite mass that can prevent the viscous spreading of such a ring is of the order of $10^{-8}$.
\item
Our preliminary experiments with self-gravity indicate that the confinement mechanisms works also for self-gravitating particles, although due to the enhanced viscosity, the threshold $\mu \propto \sqrt{\nu}$ is increased. At low $\tau$ the enhancement in $\nu$ is due to gravitational encounters and amount to less than a factor of 10 compared to non-gravitating values. In large $\tau$, the gravitational wakes may enhance the viscosity even by a factor of 100. The implied increase in threshold $\mu$ is thus by a factor $3-10$.

\end{enumerate}

Our results provide encouraging evidence that second-order resonances can confine ring material around a non-axisymmetric object. However, several pieces of the puzzle are still missing to fully explain this confinement. 

The initial material confinement observed at the 1/3 SOR (see the lowermost panels of Fig.~\ref{fig_map_J_e_snapshots_phases_I_II_III}) can easily be understood by the fact that the velocity field in the forming ringlet is reversed when compared to the background Keplerian motion: at its outermost radial position, a ring particle moves slower than the background particles on circular orbits, and vice-versa at its innermost position. Things get more complicated when the ring becomes narrow. The confinement is then reminiscent of the ``single sided shepherding" seen in the simulations of \cite{hanninen1994, hanninen1995} that described ring confinement at Lindblad resonances forced by a satellite. This result was explained analytically by the negative angular momentum luminosity forced by the satellite \citep{goldreich1995}. Unfortunately, such analytical model is not readily applicable to the 1/3 SOR, due to the impossibility to avoid self-intersection for a ``pure 1/3 SOR streamline". Thus, the next important step is to explain how the 1/3 SOR excitation is transferred to free Lindblad modes, and how these modes lead to the reversal of angular momentum flux.

While analytical expressions for the torques exerted by Lindblad resonances have been derived decades ago, there is no such expressions for second-order resonances, while such torque is observed in our simulation, see for instance the passage through the 3/5 SOR in Fig.~\ref{fig_passage_resonances_t_Lz} that show a clear jump of angular momentum. The calculation of this torque would be important to assess the long term stability of a 1/3 SOR ringlet. At this point, and considering the difficulty to describe how free Lindblad modes are excited at the 1/3 SOR, it is not clear if such analytical expression is obtainable.

Another important topic to study further is the effect of self-gravity on the ring confinement. Our small-N simulations have addressed mainly the influence of gravitational encounters, while a much larger $N$ would be needed to study realistic dense rings with self-gravity wakes. The gravitational accretion of particles in perturbed versus non-perturbed rings would also merit a dedicated study. In particular, the cyclic behavior of perturbed rings near the accretion boundary, potentially relevant for example to the time-dependent structure of Saturn's F-ring, should be verified by independent simulations.

Our simulations have focused on the case of the 1/3 SOR using Chariklo's parameters. It is now important to have simulations more specifically focused on Haumea and on the 1/3 and 5/7 SORs around Quaoar. 

Independent of rings around non-axisymmetric bodies, it is also important to extend this work to the confinement of material at second-order resonances caused by forming planets in circum-stellar disks or in the proto-Solar System.

\begin{acknowledgements}
This work has been supported by the French ANR project Roche, number ANR-23-CE49-0012.
\end{acknowledgements}

\bibliographystyle{aa}
\bibliography{references}

\begin{appendix}
\section{Potentials acting on the ring}
\label{app_potentials}

\subsection{Triaxial ellipsoid}

The analytical expansions and numerical values for the potential of homogeneous triaxial ellipsoids are given in  the Appendix~B of  Paper~I. We denote the total mass of the the body by $M$ and the semiaxes of the ellipsoid by $A,B,C$, with $A>B>C$. The reference radius $R_{\rm ref}$, elongation $\epsilon_{\rm elon}$ and oblateness $f$ of the body are the defined 
by\footnote{The parameters $\epsilon_{\rm elon}$ and $f$ are related to the classical harmonic coefficients through $\epsilon_{\rm elon} = 10C_{2,2}$ and $f= -(5/2)C_{2,0}= (5/2)J_2$.}
\begin{eqnarray}
 \frac{3}{{R^2_{\rm ref}}}  &=& \frac{1}{A^2}+\frac{1}{B^2}+\frac{1}{C^2}, \\ \nonumber\\
 \epsilon_{\rm elon}        &=& \frac{A^2-B^2}{2{R^2_{\rm ref}}}, \\ \nonumber\\ 
 f                          &=& \frac{A^2+B^2-2C^2}{4{R^2_{\rm ref}}}.
\end{eqnarray}

\noindent Outside the ellipsoidal body, but near to its equatorial plane
($|z|<<r$) we may approximate its potential in co-rotating
cylindrical coordinates as
\begin{eqnarray}
   U(r,\theta,z) = && \!\!\!\!\!\!\!\!\! -\frac{GM}{r} \left[ 1+ \frac{f}{5} \left( \frac{{R_{\rm ref}}}{r}\right)^2 +\sum_{p=1}^\infty
  \left( \frac{{R_{\rm ref}}}{r}\right)^{2p} S_p \epsilon_{\rm elon}^p \cos(2p \theta) \right]   \nonumber \\  \nonumber \\ 
  && \!\!\!\!\!\!\!\!\! -\frac{1}{2} \frac{GM}{r^3} \left[ 1+\frac{9f}{5}  \left( \frac{{R_{\rm ref}}}{r}\right)^2 \right] z^2.
 \label{eq_pote_ellipsoid}
\end{eqnarray}
In this formula the origin is fixed to the center of the body and the
angle $\theta$ is measured from the longest semi-axis $A$. The
method to calculate the coefficients $S_p$ is given in  the Appendix~B of  Paper~I, and
in practice we include terms up to $p=10$. 

We assume that the central body rotates counterclockwise around its
shortest axis (z-axis), so that at time $t$ the longest axis makes an
angle $\Omega_{\rm B} t$ with respect to the x-axis of the inertial system (initially
the central body long axis is aligned with the x-axis). At each force
evaluation step, the non-rotating cartesian coordinates of particles
are transformed to the system aligned with the body ellipsoid, by
rotation with the angle $\Omega_{\rm B} t$ around the z-axis.  The
accelerations components in this system are calculated from $-\nabla
U$ and then transformed back to the non-rotating system by rotation
with the angle $-\Omega_{\rm B} t$.

\subsection{Mass anomaly}

In the Appendix~A of  Paper~I, the potential caused by a point-like mass anomaly on
the central body was expanded in rotating cylindrical coordinates centered onto 
the body itself, the most convenient approach to investigate the structure of the 
various resonances at play. 

In numerical simulations, it is simpler and computationally faster to
write the forces directly in the cartesian center-of-mass
system. Let $M$ stand for the total mass of the central body
(including the mass anomaly), and define $\mu$ as the dimensionless fractional
mass contained in the mass anomaly, located at the distance $R_{\rm ref}$ from
the center of the body along its longest semi-axis. In the
center-of-mass system and at time $t$, the center of the body and the
mass anomaly have respective coordinates 
$\vec r_0 = (x_0,y_0,0)$ and $\vec r_1 = (x_1,y_1,0)$, where
\begin{eqnarray}
    x_0 = & \displaystyle -\frac{\mu}{1+\mu}R_{\rm ref} \cos(\Omega_{\rm B} t),\  y_0 = -\frac{\mu}{1+\mu}R_{\rm ref} \sin(\Omega_{\rm B} t),    \nonumber \\ 
    x_1 = & \displaystyle \ \frac{1}{1+\mu}\ R_{\rm ref} \cos(\Omega_{\rm B} t), \   y_1= \ \ \ \frac{1}{1+\mu}R_{\rm ref} \sin(\Omega_{\rm B} t).
\end{eqnarray}

\noindent The acceleration felt by a ring particle $i$ is
\begin{eqnarray}
  \ddot{\vec r}_i &=&- GM(1-\mu) \frac{\vec r_{i0}}{{r_{i0}}^3} -GM\mu \frac{\vec r_{i1}}{{r_{i1}}^3} \nonumber \\ \nonumber \\
  &=& \ \ - \nabla_i \left[ -  \frac{GM(1-\mu)}{|\vec r_{i0}|} - \frac{GM\mu}{|\vec r_{i1}|}\right],
   \label{eq_pote_mu}
\end{eqnarray}

\noindent where the abbreviation $\vec r_{ij} = \vec r_i -\vec r_j$ is used.
 
In case iii) where the mass anomaly is combined with a triaxial central body
shape, the first point-mass potential term in Eq.~\ref{eq_pote_mu} is replaced
by that calculated from Eq.~\ref{eq_pote_ellipsoid}.

\subsection{Jacobi constant}

Since the central body potential $U$ is time-independent in the frame rotating with constant angular velocity $\Omega_{\rm B} \hat z$, the equations of the motion conserve the Jacobi energy,
 \begin{eqnarray}
 E_{\rm J}  &=& E_{\rm kin} + U + \Omega_{\rm B} L_z   \\
      &=& \frac{1}{2} ({\dot x}^2+{\dot y}^2+{\dot z}^2)+U(x,y,z) - \Omega_{\rm B} (x\dot y -y \dot x),
      \label{eq_ej}
 \end{eqnarray}
where $E_{\rm kin}$ and $L_z$ denote the kinetic energy and angular momentum z-component in the center-of mass inertial frame. This is a useful quantity that can be used for checking the accuracy of the numerical integrations, in case impacts are ignored.

\subsection{Satellites}

In the case when an additional satellite with mass $M_{\rm s}$ is included, the
integrations are also performed in the system centered at the mass-center of
the system formed by the central body and its mass anomaly. In this system, the
acceleration of the satellite located at $\vec r_s$ is obtained from
Eq.~\ref{eq_pote_mu}, except that the total mass $M_{\rm s}+M$ replaces $M$
(see e.g. \citealt{valtonen2006}, p. 31).

\begin{eqnarray}
 \ddot{\vec r}_s =  - G(M+M_{\rm s})(1-\mu) \frac{\vec r_{\rm s0}}{{r_{\rm s0}}^3} -G(M+M_{\rm s})\mu \frac{\vec r_{\rm s1}}{{r_{\rm s1}}^3}.
  \label{eq_pote_sat}
\end{eqnarray}
For the ring particles, the perturbing acceleration (added to  Eq.~\ref{eq_pote_mu}) due
satellite includes both the direct and indirect parts:

\begin{eqnarray}
 \Delta  \ddot{\vec r}_i =&-& GM_{\rm s}\frac{\vec r_{is}}{{r_{is}}^3} \nonumber \\
  &-& GM_{\rm s} (1-\mu) \frac{\vec r_{\rm s0}}{{r_{\rm s0}}^3} -GM_{\rm s}\mu \frac{\vec r_{\rm s1}}{{r_{\rm s1}}^3}.
\end{eqnarray}

\section{Treatment of impacts}
\label{app_treatment_impacts}

For particle impacts we employ the visco-elastic "shock absorber" technique, initially introduced by \cite{salo1995} for
self-gravitating local simulations pertaining to Saturn's rings. This method involves integrating the motion of particle pairs through each
collision event. During the impact the particle pairs experience a repulsive force that is directly proportional to the extent of their
mutual overlap. Additionally, they encounter a viscous force proportional to the perpendicular component of their relative
velocity. The combination of these force components ensures that the colliding particles rebound with a
decreased post-collisional relative speed.  The two parameters contained in this linear force model can be written in terms of the
desired coefficient of restitution $\epsilon_{\rm n} = |{v_{\rm n}}^\prime|/|{v_{\rm n}}|$, which specifies the ratio of post and
pre-collisional velocity differences, and the duration of the impact $T_{\rm dur}$ (see \citealt{salo2018}).  In all our current simulations
we employ $\epsilon_{\rm n}=0.1$.

The advantage of such ``soft-particle" force-method, compared to the more-standard treatment of impacts as instantaneous velocity changes,
is its ability to handle multiple simultaneous impacts and gravitational sticking of particles, without leading to artificial
overlaps of particles. In addition, the technique allows for realistic inclusion of ring self-gravity, even when self-gravity is strong enough
to lead to gravitationally bound particle groups (see \citealt{salo1995,mondino2022}). Recently, \cite{torii2024, torii2025} adopted the same method in their global simulations of embedded satellites in Saturn's rings. 

The downside of the soft-particle method is that impacts need to be resolved with very small time steps, which increases the CPU consumption. To alleviate this problem, we use two tricks. The first is to scale-up the impact duration to be longer than the actual physical impact duration (a fraction of second), but still keep $T_{\rm dur}$ short compared to orbital period $T_{\rm orb}$: tests reported in \cite{salo2018} indicate that impact durations of the order of $T_{\rm dur}= 0.0025 T_{\rm orb}$ yield results that are indistinguishable from treating impacts as instantaneous. { Similar result was found by \cite{torii2025}}.  For too long impact duration, the steady-state velocity dispersion following from the balance between viscous gain and collisional dissipation would be modified. For example for the Chariklo ringlets, $T_{\rm dur}= 0.0025 T_{\rm orb}$ corresponds to $\sim 100$ seconds.  Another trick to speed up the calculations is to use a separate treatment of those particles which are currently colliding, compared to those which are not.  The short time  steps $\Delta t_{\rm small} << T_{\rm dur}$,  dictated by the duration of impact, are only used for the integration of currently colliding pairs, while the other particles are integrated with longer time steps $\Delta t << T_{\rm orb}$, whose length is determined by the need to integrate accurately the orbital motion.

Typically, we choose the constants of the visco-elastic force model in such a way that $T_{\rm dur} \sim 0.0015 T_{\rm orb}$, and employ an integration time step $\Delta t_{\rm small} = 0.1 T_{\rm dur}$ for particles currently colliding. With these values, the maximum penetration in impacts typically stays below $1\%$ of particle radius\footnote{Except for the transient resonance excitation phase where velocity dispersion is very high.}.  For orbital integration, we use $\Delta t =\frac{1}{300} T_{\rm orb}$ unless otherwise indicated.  In the self-gravitating simulation of Section~\ref{sec_sg}, two-times smaller $T_{\rm dur}$, $\Delta t$, and $\Delta t_{\rm small}$ were used.

In practice it is important to keep the number of pairs requiring small time-steps $\Delta t_{\rm small}$ as low as possible.  To achieve this we construct in the beginning of each orbital integration step  a list of particle pairs that may collide during the next large time step $\Delta t$.  This list is constructed in two stages: we first make an initial list of all particle pairs whose mutual distances are less than a given threshold distance $R_{\rm lim1}$. Search of these close pairs makes use of the fact that the system is a narrow annulus and orders the particles according to their azimuthal coordinate. In the second stage, we utilize particle velocities and use a first order Taylor-expansion to estimate the minimum separation each pair included to the initial list can achieve during the orbital integration step  $\Delta t$.  If this minimum separation is smaller than a threshold separation $R_{\rm lim2}$, then the pair is accepted to a refined list of potentially colliding pairs.

In the calculation of forces on particles in RK4, all particles which are members of the refined list of pairs (``potential impactors"), are integrated with the small time step $\Delta t_{\rm small}$. Not all of them are currently experiencing an impact: at each small time step, we check whether the pair actually overlaps and if so, then impact forces are added.  On the other hand, those particles which are not in the list of potential impactors, are integrated with the large time step $\Delta t$.  At low optical depth systems we concentrate on, only a small fraction of particles are impacting at a given instant of time, and therefore such a splitting of particles to impacting and non-impacting particles gives easily an order of magnitude speed-up. In practice, it is important to keep the list of potential impactors as short as possible, while still making sure that all actual impacts are found. This is achieved by dynamically adjusting both $R_{\rm lim1}$ and $R_{\rm lim2}$. The first threshold is based on the velocity dispersion of the system, and the second one is based on what has been, during the last few integrations steps,  the maximum pre-step separation actually leading to an impact.

In our self-gravitating simulations, the additional forces between particles were calculated after every $\Delta t$, together with the time derivatives of the forces. Due to small $N$ in the current simulations, a particle-particle force calculation method was used, including all particle pairs within a certain limiting distance $R_{\rm grav}$. This limiting distance was chosen to be of the order of 100 particle diameters and at least 10 times bigger than the largest aggregate forming in the simulation. In practice the calculation of forces was done together with finding the nearby impact pairs. Since $R_{\rm grav} >> R_{\rm lim1}$,  the CPU-time consumption in finding the pairs was substantially larger in self-gravitating simulations compared to non-gravitating simulations, in particular if large aggregates formed. During the small integration time steps, a linear Taylor approximation of forces was used: in \cite{karjalainen2004} it was found that such an approximation improves significantly the accuracy of the gravity calculation. In particular, it removed the secular rotational instability of the aggregates seen in case constant forces were used during the step $\Delta t$.

\section{Additional simulations}
\label{AppendixC}
In this section, we present complementary simulations to test some of the results obtained in the main text.

\subsection{Effect of nearby first-order 2/3 resonance on 1/3 SOR}
\label{AppendixC1}
\begin{figure*}[!ht]
  \centering
 \includegraphics[width=1.8\columnwidth]{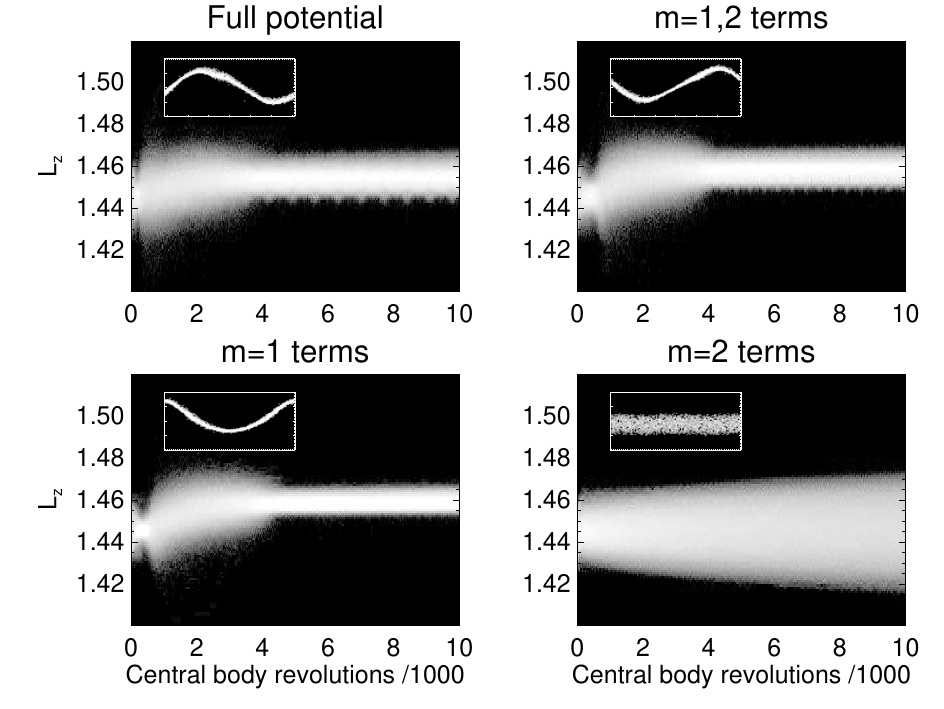}
\caption{Four collisional simulations of 1/3 SOR with $\mu=0.1$, $R=0.001$ (corresponds to 200~m for Chariklo's rings).
\textit{Upper row, left}: standard simulation using full mass anomaly potential. \textit{Upper row, right}: using an expansion of the potential,
including m=1 and m=2 terms.
\textit{Lower row}: using the $m=1$ (left) or $m=2$ (right) terms of the potential expansion.
}
\label{fig_pot_modes}
\end{figure*}

Figure~\ref{fig_resonance_overlap} indicated that the theoretical eccentricity amplitudes excited by the first-order 2/3 resonance are very large at the vicinity of the 1/3 SOR, even 30\% for $\mu=0.1$. To demonstrate that this resonance overlap does not affect the confinement seen at collisional 1/3 resonance simulations, we have conducted additional simulations, where only the $m=2$ and $m=1$ terms of the mass anomaly potential are taken into account. As seen in Fig.~\ref{fig_pot_modes}, the $m=2$ term is not able to lead to any secular change in $L_z$ distribution at the 1/3 SOR. On the other hand, using just the $m=1$ leads to essentially similar evolution as using the full potential.

\subsection{Wide initial distribution}
\label{AppendixC2}

\begin{figure*}[!ht]
\centering
 \includegraphics[width=1.6\columnwidth]{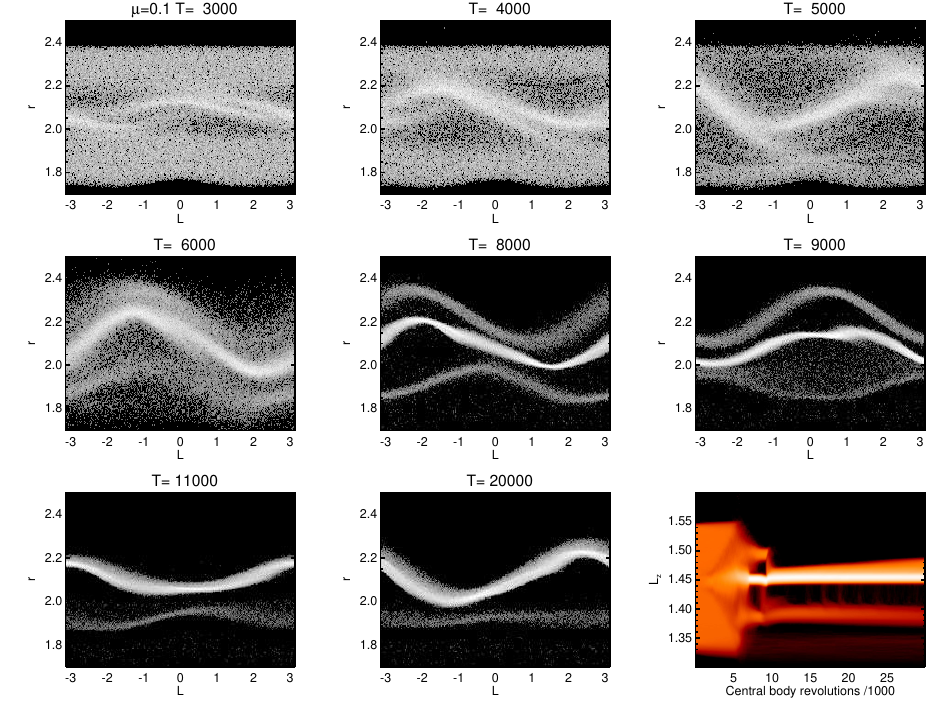}
\caption{
 The time evolution of a ring perturbed by a large mass anomaly $\mu=0.1$, to be compared with Fig.~\ref{fig_map_J_e_snapshots_phases_I_II_III}.
 The ring is initially centered around 1/3 resonance distance $a_0$ with a full width of 0.6. This is slightly larger that the theoretical maximum of epicyclic excursions, $a_0 e_{\rm peak} \approx 0.27$ and over a factor of ten larger than the width of the resonance zone, $W_{\rm res} \approx 0.026$. Particle size $R=0.001$ corresponds to 200 m for Chariklo's rings. The three upper panels show the self-intersecting streamline forced by the 1/3 SOR. This self-crossing disappears around $T=$~6,000 to form a streamline dominated by a $m=1$ azimuthal mode. In contrast to the case of narrower initial rings (Fig.~\ref{fig_map_J_e_snapshots_phases_I_II_III}), two supplementary rings form both interior and exterior to the main ring at $T=$~6,000-8,000. However, the outer ringlet is rapidly swallowed by the main ringlet (at $T\approx$~10,000), while the inner ringlet survives until the end of the simulation. The last panel at lower right  (red color palette) shows the time evolution of the $L_z$ distribution until the end of the simulation at $T=$~30,000.
}
\label{fig_wide_initial_distribution}
\end{figure*}

Figure~\ref{fig_map_J_e_snapshots_phases_I_II_III} shows the confinement of a ring near the 1/3 SOR using a small mass anomaly $\mu=0.003$. In order to better show the formation of the initial self-intersecting streamline forced by the resonance, we have performed a simulation with a large mass anomaly $\mu=0.1$, see Fig.~\ref{fig_wide_initial_distribution}. It clearly shows the formation of a self-intersecting streamline forced by the 1/3 SOR and its transformation into a single ringlet. Due to large initial width of the particle distribution, too additional ringlets accumulate in the same resonance site, the outer of them eventually merging with the central feature.

\subsection{Additional tests of confinement criterion}
\label{AppendixC3}

\begin{figure*}[ht]
\centering
  \includegraphics[width=1.8\columnwidth]{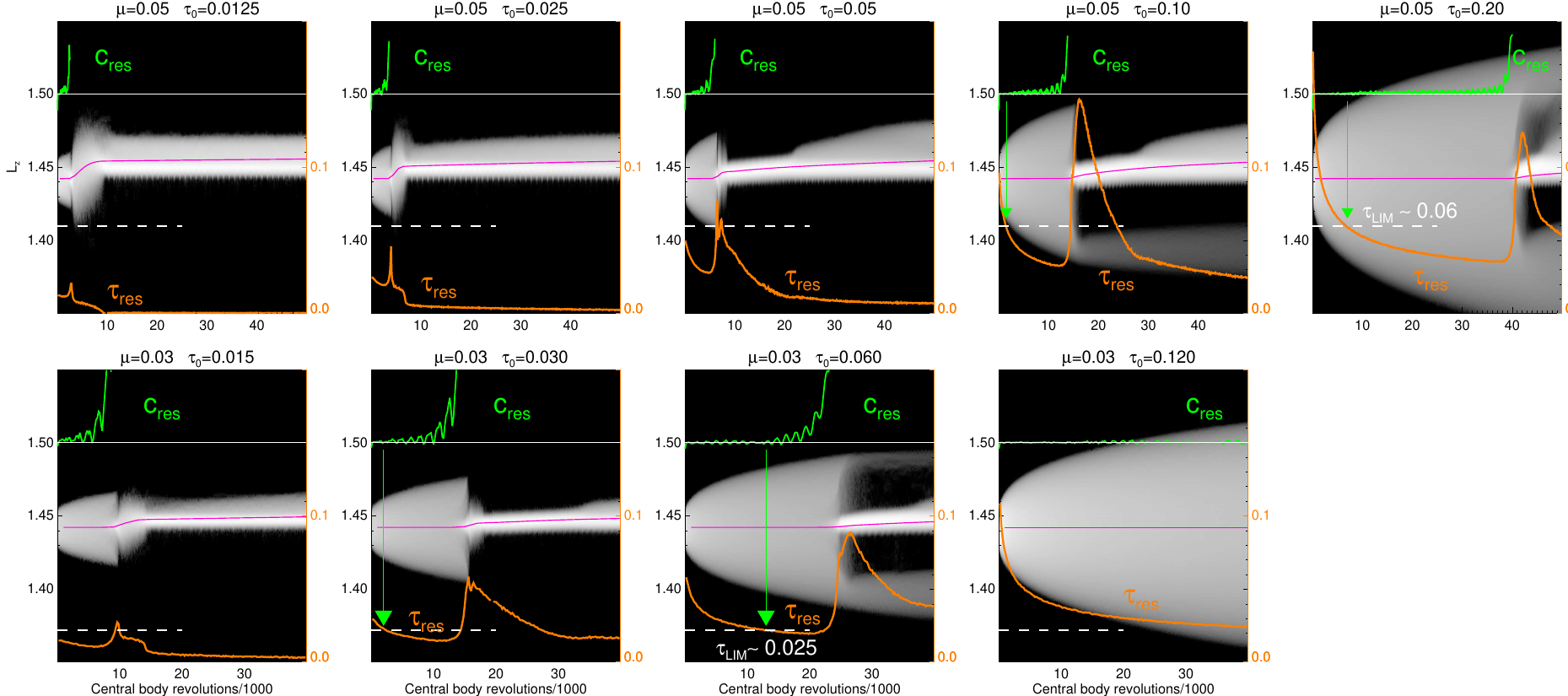}
  \caption{
  The effect of optical depth on the resonance accumulation.
  The time evolution of the $L_z$ distribution is shown with various initial optical depths $\tau_0$ and a fixed particle radius $R=0.001$ (corresponds to 200 m for Chariklo's rings). Upper row: mass anomaly of $\mu=0.05$; lower row: $\mu=0.03$. Because $R$ is fixed, different $\tau_0$'s correspond to different numbers of particles.  The orange curve shows the evolution of the optical depth at the resonance zone, calculated from the number of particles with $|a-a_0| < W_{\rm res}$ (axis labels in the right), while the green curve indicates the excess radial velocity dispersion at the same region (full scale corresponds to 5-fold dispersion compared to non-perturbed dispersion). Dashed lines indicate the estimated $\tau_{\rm lim}$: for $\tau_{\rm res} < \tau_{\rm lim}$ the excitation of resonance perturbations (indicated by the growing oscillations of $c_{\rm res}$) starts, either initially in case $\tau_0 \lesssim t_{\rm lim}$, or after a sufficient drop of $\tau_{\rm res}$ caused by the initial viscous diffusion. The latter cases are indicated by green arrows and provide estimate of $\tau_{\rm lim}$. Based on the simulations $\tau_{\rm lim} \approx 0.06$ and  $0.025$ for $\mu=0.05$ and $0.03$, respectively.}
  \label{fig_effect_tau_on_dispersal}
\end{figure*}

The Fig.~\ref{fig_threshold} explored the effects of various parameters of the runs (particle radius $R$ and mass anomaly $\mu$) in order to test Eq.~\ref{eq_threshold}. The Fig.~\ref{fig_effect_tau_on_dispersal} complements this testing by examining more specifically the effect of $\tau$ on the ring confinement for a given particle radius of $R=0.001$.

According to Eq.~\ref{eq_threshold},  $\tau_{\rm lim}$ is proportional to $\mu^2$ for a fixed $R$. To test this behavior, two additional series of simulations were performed with varying $\tau_0$, with $\mu=0.05$ and $0.03$ (Fig.~\ref{fig_effect_tau_on_dispersal}). According to  Eq.~\ref{eq_threshold} the expected $\tau_{\rm lim} \approx 0.1$ and $0.035$, respectively, but based on simulations the actual limits are about 0.06 and 0.025, respectively. 
Note that in the case of $\mu=0.05$, all the simulations with $\tau_0$ up to 0.2 eventually lead to accumulation at the resonance zone. However, this does not contradict the above threshold criterion, since the optical depth at the resonance zone drops during the simulation due to viscous spreading. See the caption of Fig.~\ref{fig_effect_tau_on_dispersal}.

\subsection{Shear reversal in first-order resonance}
\label{AppendixC4}
\begin{figure*}
  \centering
  \includegraphics[width=2\columnwidth] {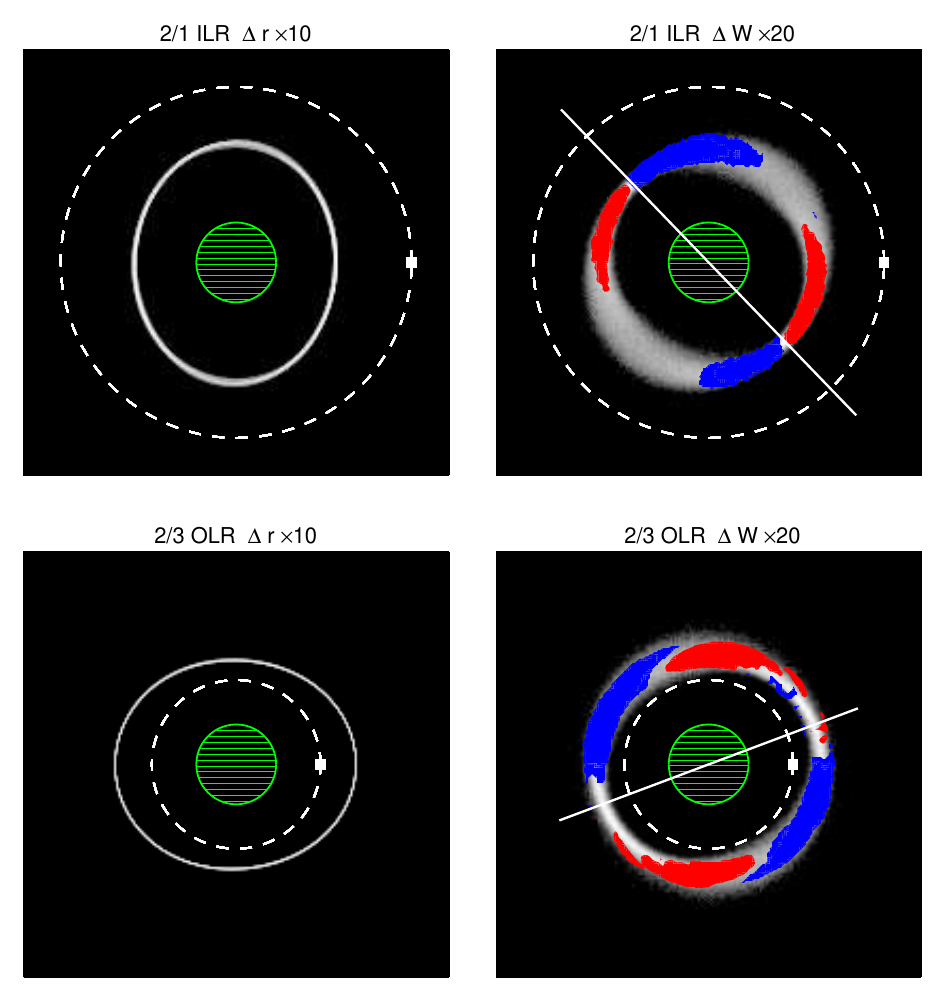}
  \caption{
  Forced response of a collisional ring to first-order Lindblad resonances:  2/1 inner resonance $(m=2, j=1)$ in the upper row and 2/3 outer resonance $(m=-2, j=1)$ in the lower row. In both cases the ringlet is centered at unit distance, and is composed of 7,500 particles with radius 0.0005; the satellite has a mass $\mu_{\rm s} =2 \times 10^{-4}$ relative to the spherical central body. In the frames 50 snapshots have been stacked in a coordinate system corotating with the satellite (the satellite location is shown by the white square). In the left frame, the particle deviations from the ringlet mean distance have been exaggerated by a factor of 10. In the right frames the ringlet width has been exaggerated by a factor of 20. In right frames the blue (red) contours indicate regions where $T_{rt}$ is negative (positive). The line indicates the direction where the ring width is the smallest. Local shear reversal, corresponding to negative $T_{rt}$, is seen in the ring regions where particles are approaching the width minimum.
  }
  \label{fig_ilr_olr}
  \end{figure*}

Figure~\ref{fig_ilr_olr} shows examples of the response of collisional ring to first-order 2/1 and 2/3 satellite resonances, in a frame rotating with the satellite. In case of 2/1, the $m=2$ response is oriented nearly perpendicular to the instantaneous direction of the satellite, while for 2/3 the orientation is nearly aligned with the satellite. Due to impacts, there is a slight offset in the response (seen best in the right frames, emphasizing the ring width variations), leading to negative (positive) torque exerted on the ring by the outer (inner) satellite.  Also shown in the right frames are the regions where local shear is reversed: blue regions indicate where $T_{rt}=<\Delta v_r \Delta v_t>$ is negative, $\Delta v_r$ and $\Delta v_t$ denoting the particles' radial and tangential velocities compared to the mean flow velocity at its location.

\end{appendix}

\end{document}